\documentclass[aps,eqsecnum,preprint,floats,epsf,epsfig,nofootinbib]{revtex4}
\usepackage{graphicx}
\usepackage{epsfig}

\begin{document}
\def\be{\begin{eqnarray}}
\def\en{\end{eqnarray}}
\def\non{\nonumber}
\def\la{\langle}
\def\ra{\rangle}
\def\pp{{\prime\prime}}
\def\nc{N_c^{\rm eff}}
\def\vp{\varepsilon}
\def\hep{\hat{\varepsilon}}
\def\drho{\bar\rho}
\def\deta{\bar\eta}
\def\a{{\cal A}}
\def\B{{\cal B}}
\def\c{{\cal C}}
\def\d{{\cal D}}
\def\e{{\cal E}}
\def\p{{\cal P}}
\def\t{{\cal T}}
\def\B{{\cal B}}
\def\L{{\cal L}}
\def\P{{\cal P}}
\def\S{{\cal S}}
\def\T{{\cal T}}
\def\C{{\cal C}}
\def\A{{\cal A}}
\def\E{{\cal E}}
\def\V{{\cal V}}
\def\CP{$CP$~}
\def\up{\uparrow}
\def\dw{\downarrow}
\def\vma{{_{V-A}}}
\def\vpa{{_{V+A}}}
\def\smp{{_{S-P}}}
\def\spp{{_{S+P}}}
\def\lrpartial{\buildrel\leftrightarrow\over\partial}
\def\J{{J/\psi}}
\def\3bar{{\bf \bar 3}}
\def\6bar{{\bf \bar 6}}
\def\10bar{{\bf \ov{10}}}
\def\ov{\overline}
\def\Lqcd{{\Lambda_{\rm QCD}}}
\def\pr{{Phys. Rev.}~}
\def\prl{{ Phys. Rev. Lett.}~}
\def\pl{{ Phys. Lett.}~}
\def\np{{ Nucl. Phys.}~}
\def\zp{{ Z. Phys.}~}
\def\lsim{ {\ \lower-1.2pt\vbox{\hbox{\rlap{$<$}\lower5pt\vbox{\hbox{$\sim$}
}}}\ } }
\def\gsim{ {\ \lower-1.2pt\vbox{\hbox{\rlap{$>$}\lower5pt\vbox{\hbox{$\sim$}
}}}\ } }

\font\el=cmbx10 scaled \magstep2{\obeylines\hfill BNL-HET-05/7}

\font\el=cmbx10 scaled \magstep2{\obeylines\hfill May, 2005}

\vskip 1.5 cm

\centerline{\large\bf Effects of Final-state Interactions on
Mixing-induced $CP$ Violation}
 \centerline{\large\bf in
Penguin-dominated $B$ Decays }
\bigskip
\centerline{\bf Hai-Yang Cheng,$^1$ Chun-Khiang Chua$^1$ and
Amarjit Soni$^2$}
\medskip
\centerline{$^1$ Institute of Physics, Academia Sinica}
\centerline{Taipei, Taiwan 115, Republic of China}
\medskip

\medskip
\centerline{$^2$ Physics Department, Brookhaven National
Laboratory} \centerline{Upton, New York 11973}
\medskip

\bigskip
\bigskip
\centerline{\bf Abstract}
\bigskip
\small
 Motivated by the recent indications of the possibility of sizable
deviations of the mixing-induced \CP violation parameter, $S_f$,
in the penguin-dominated $b\to sq\bar q$ transition decays such as
$B^0\to (\phi,\omega,\rho^0,\eta',\eta,\pi^0,f_0)K_S$ from $\sin
2\beta$ determined from $B\to J/\psi K_S$, we study final-state
rescattering effects on their decay rates and \CP violation.
Recent observations of large direct \CP asymmetry in modes such as
$B^0\to K^+\pi^-,\rho^-\pi^+$ means that final state phases in
2-body $B$ decays may not be small. It is therefore important to
examine these long-distance effects on penguin-dominated decays.
Such long-distance effects on $S_f$ are found to be generally
small [i.e. ${\cal O}(1-2\%)$] or negligible except for the
$\omega K_S$ and $\rho^0K_S$ modes where $S_f$ is lowered by
around 15\% for the former and increased by the same percentage
for the latter. However, final-state rescattering can enhance the
$\omega K_S,~\phi K_S,~\eta'K_S,~\rho^0K_S$ and $\pi^0K_S$ rates
significantly and flip the signs of direct \CP asymmetries of the
last two modes. Direct \CP asymmetries in $\omega K_S$ and
$\rho^0K_S$ channels are predicted to be $\A_{\omega K_S}\approx
-0.13$ and $\A_{\rho^0K_S}\approx 0.47$, respectively. However,
direct \CP asymmetry in all the other $b\to s$ penguin dominated
modes that we study is found to be rather small ($\lsim$ a few
percents), rendering these modes a viable place to search for the
$CP$-odd phases beyond the standard model. Since $\Delta
S_f(\equiv -\eta_fS_f-S_{J/\psi K_S}$, with $\eta_f$ being the \CP
eigenvalue of the final state $f$) and $\A_f$ are closely related,
the theoretical uncertainties in the mixing induced parameter
$S_f$ and the direct \CP asymmetry parameter $\A_f$ are also
coupled. Based on this work, it seems difficult to accommodate
$|\Delta S_f| > 0.10$ within the SM for
$B^0\to(\phi,\omega,\rho^0,\eta',\eta,\pi^0)K_S$, in particular,
$\eta'K_S$ is especially clean in our picture. For $f_0K_S$,  at
present we cannot make reliable estimates. The sign of the central
value of $\Delta S_f$ for all the modes we study is positive but
quite small, compared to the theoretical uncertainties, so that at
present conclusive statements on the sign are difficult to make.

\eject
\section{Introduction and Motivation}

Possible New Physics beyond the Standard Model is being
intensively searched via the measurements of time-dependent \CP
asymmetries in neutral $B$ meson decays into final \CP eigenstates
defined by
 \be
 {\Gamma(\ov B(t)\to f)-\Gamma(B(t)\to f)\over
 \Gamma(\ov B(t)\to f)+\Gamma(B(t)\to
 f)}=\S_f\sin(\Delta mt)+\A_f\cos(\Delta mt),
 \en
where $\Delta m$ is the mass difference of the two neutral $B$
eigenstates, $S_f$ monitors mixing-induced \CP asymmetry and
$\A_f$ measures direct \CP violation (in terms of the BaBar
notation, $\C_f=-\A_f$). The $CP$-violating parameters $\A_f$ and
$\S_f$ can be expressed as
 \be
 \A_f=-{1-|\lambda_f|^2\over 1+|\lambda_f|^2}, \qquad \S_f={2\,{\rm
 Im}\lambda_f\over 1+|\lambda_f|^2},
 \en
where
 \be
 \lambda_f={q_B\over p_B}\,{A(\ov B^0\to f)\over A(B^0\to f)}.
 \en
In the standard model $\lambda_f\approx \eta_f e^{-2i\beta}$ [see
Eq. (\ref{eq:Sf}) below] for $b\to s$ penguin-dominated or pure
penguin modes with $\eta_f=1$ ($-1$) for final $CP$-even (odd)
states. Therefore, it is expected in the Standard Model that
$-\eta_fS_f\approx \sin 2\beta$ and $\A_f\approx 0$ with $\beta$
being one of the angles of the unitarity triangle.

The mixing-induced $CP$ violation in $B$ decays has been already
observed in the golden mode $B^0\to J/\psi K_S$ for several years.
The current average of BaBar \cite{BaBarJpsi} and Belle
\cite{BelleJpsi} measurements is
 \be
 \sin 2\beta\approx S_{J/\psi K_S}=0.726\pm0.037\,.
 \en
However, the time-dependent {\it CP}-asymmetries in the $b\to
sq\bar q$ induced two-body decays such as $B^0\to
(\phi,\omega,\pi^0,\eta',f_0)K_S$ are found to show some
indications of deviations from the expectation of the Standard
Model (SM). The BaBar \cite{BaBarSf} and Belle \cite{BelleSf}
results and their averages are shown in Table \ref{tab:Data}. In
the SM, \CP asymmetry in all above-mentioned modes should be equal
to $S_{J/\psi K}$ with a small deviation {\it at most} ${\cal
O}(0.1)$ \cite{LS}. As discussed in \cite{LS}, this may originate
from the ${\cal O}(\lambda^2)$ truncation and from the subdominant
(color-suppressed) tree contribution to these processes. From
Table \ref{tab:Data} we see some possibly sizable deviations from
the SM, especially in the $\eta' K_S$ mode in which the
discrepancy $\Delta S_{\eta'K_S}=-0.30\pm0.11$ is a $2.7\sigma$
effect where
 \be
 \Delta S_f\equiv -\eta_fS_f-S_{J/\psi K_S}.
 \en
If this deviation from $S_{J/\psi K}$ is confirmed and established
in the future, it may imply some New Physics beyond the SM.

\begin{table}[h]
\caption{Mixing-induced and direct \CP asymmetries $-\eta_f S_f$
(first entry) and $\A_f$ (second entry), respectively, for various
penguin-dominated modes with $\eta_f$ being the \CP eigenvalue of
the final state. Experimental results are taken from
\cite{BaBarSf,BelleSf}. } \label{tab:Data}
\begin{ruledtabular}
\begin{tabular}{c r r r}
Final State & BaBar \cite{BaBarSf} & Belle \cite{BelleSf} & Average  \\
\hline
 $\phi K_S$ & $0.50\pm0.25^{+0.07}_{-0.04}$ & $0.08\pm0.33\pm0.09$
 & $0.35\pm0.20$ \\
 & $-0.00\pm0.23\pm0.05$ & $0.08\pm0.22\pm0.09$ & $0.04\pm0.17$ \\
 $\omega K_S$ & $0.50^{+0.34}_{-0.38}\pm0.02$ & $0.76\pm0.65^{+0.13}_{-0.16}$ & $0.55^{+0.30}_{-0.32}$ \\
 & $0.56^{+0.29}_{-0.27}\pm0.03$ & $0.27\pm0.48\pm0.15$ & $0.48\pm0.25$ \\
 $\eta'K_S$ & $0.30\pm0.14\pm0.02$ & $0.65\pm0.18\pm0.04$ &
 $0.43\pm0.11$ \\
 & $0.21\pm0.10\pm0.02$ & $-0.19\pm0.11\pm0.05$ & $0.04\pm0.08$ \\
 $\pi^0K_S$ & $0.35^{+0.30}_{-0.33}\pm0.04$ & $0.32\pm0.61\pm0.13$
 & $0.34^{+0.27}_{-0.29}$ \\
 & $-0.06\pm0.18\pm0.03$ & $-0.11\pm0.20\pm0.09$ & $-0.08\pm0.14$ \\
 $f_0K_S$ & $0.95^{+0.23}_{-0.32}\pm0.10$ & $-0.47\pm0.41\pm0.08$
 & $0.39\pm0.26$ \\
 & $0.24\pm0.31\pm0.15$ & $-0.39\pm0.27\pm0.09$ & $-0.14\pm0.22$ \\
\end{tabular}
\end{ruledtabular}
\end{table}

In order to detect the signal of New Physics unambiguously in the
penguin $b\to s$ modes, it is of great importance to examine how
much of the deviation of $S_f$ from $S_{J/\psi K}$ is allowed in
the SM
\cite{LS,Grossman97,Grossman98,Grossman03,Gronau-piK,Gronau-eta'K}.
In all these previous studies and estimates of $\Delta S_f$,
effects of FSI were not taken into account. In view of the
striking observation of large direct \CP violation in $B^0\to
K^\pm\pi^\mp$, it is clear that final-state phases in two-body $B$
decays may not be small. It is therefore important to understand
their effects on $\Delta S_f$.

The decay amplitude for the pure penguin or penguin-dominated
charmless $B$ decay in general has the form
 \be
 M(\ov B^0\to f) = V_{ub}V_{us}^*F^u+V_{cb}V_{cs}^* F^c
 +V_{tb}V_{ts}^*F^t.
 \en
Unitarity of the CKM matrix elements leads to
 \be
 M(\ov B^0\to f) = V_{ub}V_{us}^*A_f^u+V_{cb}V_{cs}^*A_f^c
 \approx A\lambda^4R_be^{-i\gamma}A_f^u+A\lambda^2 A_f^c,
 \en
where $A_f^u=F^u-F^t$, $A_f^c=F^c-F^t$,
$R_b\equiv|V_{ud}V_{ub}/(V_{cd}V_{cb})|=\sqrt{\bar\rho^2+\bar\eta^2}$
with $\bar\rho,\bar\eta,A,\lambda$ being the Wolfenstein
parameters \cite{Wolfenstein}. The first term is suppressed by a
factor of $\lambda^2$ relative to the second term. For a pure
penguin decay such as $B^0\to\phi K^0$, it is naively expected
that $A_f^u$ is in general comparable to $A_f^c$ in magnitude.
Therefore, to a good approximation $-\eta_fS_f\approx\sin
2\beta\approx S_{J/\psi K}$. For penguin-dominated modes such as
$\omega K_S,\rho^0K_S,\pi^0K_S$, $A_f^u$ also receives tree
contributions from the $b\to u\bar u s$ tree operators. Since the
Wilson coefficient for the penguin operator is smaller than the
one for the tree operator, $A_f^u$ could be significantly larger
than $A_f^c$. As the first term carries a weak phase $\gamma$, it
is possible that $S_f$ is subject to a significant ``tree
pollution". To quantify the deviation, it is known that to the
first order in $r_f\equiv(\lambda_u A_f^u)/(\lambda_c A_f^c)$
\cite{Gronau,Grossman03}
 \be \label{eq:CfSf}
 \Delta S_f=2|r_f|\cos 2\beta\sin\gamma\cos\delta_f,
 \qquad \A_f=2|r_f|\sin\gamma\sin\delta_f,
 \en
with $\delta_f={\rm arg}(A_f^u/A_f^c)$. Hence, the magnitude of
the \CP asymmetry difference $\Delta S_f$ and direct \CP violation
are both governed by the size of $A_f^u/A_f^c$. However, for the
aforementioned penguin-dominated modes, the tree contribution is
color suppressed and hence in practice the deviation of $S_f$ is
expected to be small \cite{LS}. (However, we shall see below that
a sizable $\Delta S_f$ can occur in $\omega K_S$ and $\rho^0K_S$
modes. For a review of model calculations of $\Delta S_f$, see
\cite{Mishima}.)

Since the penguin loop contributions are sensitive to high
virtuality, New Physics beyond the SM may contribute to $S_f$
through the heavy particles in the loops (for a review of the New
Physics sources contributing to $S_f$, see \cite{Browder}).
Another possibility is that final-state interactions are the
possible tree pollution sources to $S_f$. Both $A_f^u$ and $A_f^c$
will receive long-distance tree and penguin contributions from
rescattering of some intermediate states. In particular there may
be some dynamical enhancement of light $u$-quark loop. If tree
contributions to $A^u_f$ are sizable, then final-state
rescattering will have the potential of pushing $S_f$ away from
the naive expectation. Take the penguin-dominated decay $\ov
B^0\to \omega \ov K^0$ as an illustration. It can proceed through
the weak decay $\ov B^0\to K^{*-}\pi^+$ followed by the
rescattering $K^{*-}\pi^+\to \omega \ov K^0$. The tree
contribution to $\ov B^0\to K^{*-}\pi^+$, which is color allowed,
turns out to be comparable to the penguin one because of the
absence of the chiral enhancement characterized by the $a_6$
penguin term. Consequently, even within the framework of the SM,
final-state rescattering may provide a mechanism of tree pollution
to $S_f$. By the same token, we note that although $\ov B^0\to
\phi \ov K^0$ is a pure penguin process at short distances, it
does receive tree contributions via long-distance rescattering.

In this work, we shall study the effects of final-state
interactions (FSI) on the time-dependent \CP asymmetries $S_f$ and
$\A_f$. In \cite{CCS} we have studied  the final-state
rescattering effects on the hadronic $B$ decays and examined their
impact on direct \CP violation. The direct $CP$-violating partial
rate asymmetries in charmless $B$ decays to $\pi\pi/\pi K$ and
$\rho\pi$ are significantly affected by final-state rescattering
and their signs are generally different from that predicted by the
short-distance approach such as QCD factorization
\cite{BBNS,BNa,BNb}. Evidence of direct \CP violation in the decay
$\ov B^0\to K^-\pi^+$ is now established, while the combined BaBar
and Belle measurements of $\ov B^0\to\rho^\pm\pi^\mp$ imply a
$3.6\sigma$ direct \CP asymmetry in the $\rho^+\pi^-$ mode
\cite{HFAG}. In fact, direct \CP asymmetries in these channels are
a lot bigger than expectations (of many people) and may be
indicative of appreciable LD rescattering effects, in general, in
$B$ decays. Our predictions for \CP violation agree with
experiment in both magnitude and sign, whereas the QCD
factorization predictions (especially for $\rho^+\pi^-$)
\cite{BNb} seem to have some difficulties with the data.

Besides some significant final-state rescattering effects on
direct \CP violation, another motivation for including FSIs is
that there are consistently 2 to 3 $\sigma$ deviations between the
central values of the QCDF predictions for penguin-dominated modes
such as $B\to K^*\pi,~K\phi,~K^*\phi,~K\eta'$ and the experimental
data \cite{BNb}. This discrepancy between theory and experiment
for branching ratios may indicate the importance of subleading
power corrections such as FSI effects and/or the annihilation
topology.

Since direct \CP violation in charmless $B$ decays can be
significantly affected by final-state rescattering, it is clearly
important to try to take into account the FSI effect on the
mixing-induced and direct \CP asymmetries $S_f$ and $\A_f$ of
these penguin dominated modes. The layout of the present paper is
as follows. In Sec. II we discuss the short-distance contributions
to the $b\to sq\bar q$ transition decays $B^0\to
(\phi,\omega,\rho^0,\pi^0,\eta',\eta,f_0)K_S$ within the framework
of QCD factorization. We then proceed to study the final-state
rescattering effects on \CP asymmetries $S_f$ and $\A_f$ in Sec.
III. Sec. IV contains our conclusions.

\section{Short-distance contributions}

\subsection{Decay amplitudes in QCD factorization}

 We shall use the QCD factorization approach
\cite{BBNS,BNa,BNb} to study the short-distance contributions to
the decays $\ov B^0\to
(\phi,\omega,\rho^0,\pi^0,\eta',\eta,f_0)\ov K^0$. In QCD
factorization, the factorization amplitudes of above-mentioned
decays are given by
 \be \label{eq:SDamp}
 \la\phi \overline K {}^0|H_{\rm eff}|\ov B {}^0\ra
  &=& \frac{G_F}{\sqrt2}(p_B\cdot\vp_\phi^*)\sum_{p=u,c}\lambda_p
  \Bigg\{
  \bigg[(a_3+a_5)(\ov K^0\phi)+(a_4^p+r_\chi^\phi a_6^p)(\ov K^0\phi)-\frac{1}{2}
  (a_7+a_9)(\ov K^0\phi)\bigg]
  \non \\
  &\times & 2 f_\phi m_\phi  F_1^{BK}(m_\phi^2)+
  f_Bf_Kf_\phi\bigg[b_3(\ov
  K^0\phi)-{1\over 2}b_3^{\rm EW}(\ov K^0\phi)\bigg]{2m_\phi\over m_B^2} \Bigg\}, \non \\
  \la\omega \overline K {}^0|H_{\rm eff}|\ov B {}^0\ra
  &=& \frac{G_F}{2}(p_B\cdot\vp_\omega^*)\sum_{p=u,c}\lambda_p
  \Bigg\{
  \bigg[a_2(\ov K^0\omega)\delta^p_u+2(a_3+a_5)(\ov K^0\omega)+\frac{1}{2}
  (a_7+a_9)(\ov K^0\omega)]\bigg]
  \non \\
  &\times & 2 f_\omega m_\omega F_1^{BK}(m_\omega^2)
  +\bigg[(a_4^p-\mu_\chi^Ka_6^p)(\omega\ov K^0)-{1\over 2}(a_{10}^p-\mu_\chi^Ka_8^p)
  (\omega\ov K^0)\bigg]  \non \\
  &\times& 2 f_K m_\omega A_0^{B\omega}(m_K^2)
  +f_Bf_Kf_\omega\bigg[b_3(\omega\ov
  K^0)-{1\over 2}b_3^{\rm EW}(\omega\ov K^0)\bigg]{2m_\omega\over m_B^2}\Bigg\}, \non \\
    \la\rho^0 \overline K {}^0|H_{\rm eff}|\ov B {}^0\ra
  &=& \frac{G_F}{2}(p_B\cdot\vp_\rho^*)\sum_{p=u,c}\lambda_p
  \Bigg\{
  \bigg[a_2(\ov K^0\rho)\delta^p_u+\frac{3}{2}
  (a_7+a_9)(\ov K^0\rho)]\bigg]
  \non \\
  &\times & 2 f_\rho m_\rho F_1^{BK}(m_\rho^2)
  -\bigg[(a_4^p-\mu_\chi^Ka_6^p)(\rho\ov K^0)-{1\over 2}(a_{10}^p-\mu_\chi^Ka_8^p)
  (\rho\ov K^0)\bigg]  \non \\
  &\times& 2 f_K m_\rho A_0^{B\rho}(m_K^2)
  -f_Bf_Kf_\rho\bigg[b_3(\rho\ov
  K^0)-{1\over 2}b_3^{\rm EW}(\rho\ov K^0)\bigg]{2m_\rho\over m_B^2}\Bigg\}, \non \\
    \la\pi^0 \overline K {}^0|H_{\rm eff}|\ov B {}^0\ra
  &=& i\frac{G_F}{2}\sum_{p=u,c}\lambda_p
  \Bigg\{
  \bigg[a_2(\ov K^0\pi^0)\delta^p_u+\frac{3}{2}
  \alpha_{3,{\rm EW}}(\ov K^0\pi^0)\bigg]
  f_\pi F_0^{BK}(m_\pi^2)(m_B^2-m_K^2) \non \\
  &-& \bigg[\alpha_4^p(\pi^0\ov K^0)-{1\over 2}\alpha_{4,{\rm EW}}^p(\pi^0\ov K^0)\bigg]
  f_K F_0^{B\pi}(m_K^2)(m_B^2-m_\pi^2) \non \\
  &-& f_Bf_Kf_\pi\bigg[b_3(\pi^0\ov
  K^0)-{1\over 2}b_3^{\rm EW}(\pi^0\ov K^0)\bigg] \Bigg\} , \non \\
  \la\eta' \overline K {}^0|H_{\rm eff}|\ov B {}^0\ra
  &=& i\frac{G_F}{\sqrt{2}}\sum_{p=u,c}\lambda_p
  \Bigg\{  F_0^{BK}(m_{\eta'}^2)(m_B^2-m_K^2) \bigg\{
  {f_{\eta'}^q\over\sqrt{2}}
  \bigg[a_2(\ov K^0\eta'_q)\delta^p_u+2\alpha_3(\ov K^0\eta'_q)
  \non \\
  &+& \frac{1}{2}\alpha_{3,{\rm EW}}(\ov K^0\eta'_q)\bigg]
  + f_{\eta'}^c\bigg[ a_2(\ov K^0\eta'_c)\delta^p_c
  +\alpha_3(\ov K^0\eta'_c)\bigg]  \non \\
  &+& f_{\eta'}^s\bigg[\alpha_3(\ov K^0\eta'_s)+\alpha_4^p(\ov K^0\eta'_s)
  -{1\over 2}\alpha_{3,{\rm EW}}(\ov K^0\eta'_s)
  - {1\over 2}\alpha^p_{4,{\rm EW}}(\eta'_s\ov K^0)\bigg]\bigg\} \non \\
  &+& F_0^{B\eta'}(m_{K}^2)(m_B^2-m_{\eta'}^2){f_K\over\sqrt{2}}
  \bigg[ \alpha_4^p(\eta'\ov K^0)-{1\over 2}\alpha^p_{4,{\rm
  EW}}(\eta' \ov K^0)\bigg]    \non \\
  &+& f_Bf_K({1\over\sqrt{2}}f_{\eta'}^q+f_{\eta'}^s)\bigg[b_3(\eta'\ov
  K^0)-{1\over 2}b_3^{\rm EW}(\eta'\ov K^0)\bigg] \Bigg\}, \non \\
 \la f_0 \overline K {}^0|H_{\rm eff}|\ov B {}^0\ra
 &=&  -\frac{G_F}{\sqrt{2}}\sum_{p=u,c}\lambda_p
 \Bigg\{ \left[ (a_4^p-r_\chi^Ka_6^p)(f_0\ov K^0)
 -{1\over 2}(a_{10}^p-r_\chi^Ka_8^p)(f_0\ov K^0) \right] \non \\
 &\times& f_KF_0^{Bf_0}(m_K^2)(m_B^2-m_{f_0}^2)
 + (2a_6^p-a_8^p)(\ov K^0f_0)\bar f_s\,{m_{f_0}\over
 m_b}F_0^{BK}(m^2_{f_0})(m_B^2-m_K^2)\non \\
  &-& f_Bf_K(\bar f_s+\bar f_u)\bigg[b_3(f_0\ov
  K^0)-{1\over 2}b_3^{\rm EW}(f_0\ov K^0)\bigg] \Bigg\},
 \en
where $F_{1,0}$ and $A_0$ denote pseudoscalar and vector form
factors in the standard convention \cite{BSW}, $\lambda_p\equiv
V_{pb} V^*_{ps}$,
 \be
 \alpha_3(M_1M_2)=a_3(M_1M_2)-a_5(M_1M_2), & \quad &
 \alpha_4^p(M_1M_2)=a_4^p(M_1M_2)+r^{M_2}_\chi a_6^p(M_1M_2),  \\
 \alpha_{3,{\rm EW}}(M_1M_2)=-a_7(M_1M_2)+a_9(M_1M_2), & \quad &
 \alpha_{4,{\rm EW}}^p(M_1M_2)=a_{10}^p(M_1M_2)+r^{M_2}_\chi
 a_8^p(M_1M_2),
 \non
 \en
and
 \be
 && r_\chi^\phi(\mu)=\frac{2
 m_\phi}{m_b(\mu)}\frac{f^\bot_\phi(\mu)}{f_\phi},\qquad
 r_\chi^\pi(\mu)={2m_\pi^2\over 2m_b(\mu)m_q(\mu)}, \non \\
 && r_\chi^K(\mu)={2m_K^2\over m_b(\mu)[m_s(\mu)+m_q(\mu)]},
 \qquad r_\chi^{\eta'}(\mu)={2m_{\eta'}^2\over 2m_b(\mu)
 m_s(\mu)}\left(1-{f^q_{\eta'}\over\sqrt{2}f^s_{\eta'}}\right),
 \en
with the scale dependent transverse decay constant $f_V^\bot$
being defined as
 \be
 \la V(p,\vp^*)|\bar
 q\sigma_{\mu\nu}q'|0\ra=f_V^\bot(p_\mu\vp^*_\nu-p_\nu\vp^*_\mu).
 \en
Note that the $a_6$ penguin term appears in the decay amplitude of
$\ov B^0\to\phi \ov K^0$ owing to the non-vanishing transverse
decay constant of the $\phi$ meson. The scalar decay constant
$\bar f_q$ of $f_0(980)$ in Eq. (\ref{eq:SDamp}) is defined by
$\la f_0|\bar qq|0\ra=m_{f_0}\bar f_q$. The decay amplitude for
$\ov B^0\to\eta \ov K^0$ is obtained from $\ov B^0\to\eta' \ov
K^0$ by replacing $\eta'\to\eta$. Note that the use of nonzero
$q^2$ in the argument of form factors in Eq. (\ref{eq:SDamp})
means that some corrections quadratic in the light quark masses
are automatically incorporated.

The effective parameters $a_i^p$ with $p=u,c$ can be calculated in
the QCD factorization approach \cite{BBNS}. They are basically the
Wilson coefficients in conjunction with short-distance
nonfactorizable corrections such as vertex corrections and hard
spectator interactions. In general, they have the expressions
\cite{BBNS,BNb}
 \be \label{eq:ai}
 a_i^p(M_1M_2) &=& c_i+{c_{i\pm1}\over N_c}
  +{c_{i\pm1}\over N_c}\,{C_F\alpha_s\over
 4\pi}\Big[V_i(M_2)+{4\pi^2\over N_c}H_i(M_1M_2)\Big]+P_i^p(M_2),
 \en
where $i=1,\cdots,10$,  the upper (lower) signs apply when $i$ is
odd (even), $c_i$ are the Wilson coefficients,
$C_F=(N_c^2-1)/(2N_c)$ with $N_c=3$, $M_2$ is the emitted meson
and $M_1$ shares the same spectator quark with the $B$ meson. The
quantities $V_i(M_2)$ account for vertex corrections,
$H_i(M_1M_2)$ for hard spectator interactions with a hard gluon
exchange between the emitted meson and the spectator quark of the
$B$ meson and $P_i(M_2)$ for penguin contractions. The explicit
expressions of these quantities together with the annihilation
quantities $b_3$ and $b_3^{\rm EW}$ can be found in
\cite{BBNS,BNb}.

For $B\to\eta^{(')} K$ decay, $\eta^{(')}_q$ and $\eta^{(')}_s$ in
Eq. (\ref{eq:SDamp}) refer to the non-strange and strange quark
states, respectively, of $\eta^{(')}$. The decay constants
$f_{\eta^{(')}}^{q,s,c}$ are defined by
 \be
 \la\eta^{(')}(p)|\bar
 q\gamma_\mu\gamma_5q|0\ra=-{i\over\sqrt{2}}f_{\eta^{(')}}^qp_\mu, \quad
 \la\eta^{(')}(p)|\bar s\gamma_\mu\gamma_5s|0\ra=-if_{\eta^{(')}}^sp_\mu, \quad
  \la\eta^{(')}(p)|\bar c\gamma_\mu\gamma_5c|0\ra=-if_{\eta^{(')}}^cp_\mu,
  \non \\
 \en
with $q=u$ or $d$. Numerically, we shall use \cite{Feldmann}
 \be
 &&  f_{\eta'}^q=89\,{\rm MeV}, \qquad f_{\eta'}^s=131\,{\rm MeV},
  \qquad f_{\eta'}^c=-6.3\,{\rm MeV},  \non \\
  &&  f_{\eta}^q=108\,{\rm MeV}, \qquad f_{\eta}^s=-111\,{\rm MeV},
  \qquad f_{\eta}^c=-2.4\,{\rm MeV},
 \en
recalling that, in our convention,  $f_\pi=132$ MeV. As for the
$B\to\eta^{(')}$ form factor, we shall follow \cite{BNa} to use
the relation
$F_0^{B\eta^{(')}}=(f_{\eta^{(')}}^q/f_{\eta^{(')}}^s)F_0^{B\pi}$
to obtain the values of the form factors $F_0^{B\eta^{(')}}$. The
wave functions of the physical $\eta'$ and $\eta$ states are
related to that of the SU(3) singlet state $\eta_0$ and octet
state $\eta_8$ by
 \be
 \eta'=\eta_8\sin\phi+\eta_0\cos\phi,
\qquad \eta=\eta_8\cos\phi-\eta_0 \sin\phi,
 \en
with the mixing angle  $\phi=-(15.4\pm 1.0)^\circ$
\cite{Feldmann}.

The decay $B\to f_0(980)K$ has been discussed in detail in
\cite{CYfK}. Since the scalar meson $f_0(980)$ cannot be produced
via the vector current owing to charge conjugation invariance or
conservation of vector current, the tree contribution to $\ov
B^0\to f_0\ov K^0$ vanishes under the factorization approximation.
Just as the SU(3)-singlet $\eta_0$, $f_0(980)$ in the 2-quark
picture also contains strange and non-strange quark content
 \be
 |f_0(980)\ra = |\bar ss\ra\cos\theta+|\bar nn\ra\sin\theta,
 \en
with $\bar nn\equiv (\bar uu+\bar dd)/\sqrt{2}$. Experimental
implications for the mixing angle $\theta$ have been discussed in
detail in \cite{ChengDSP,CYfK}; it lies in the ranges of
$25^\circ<\theta<40^\circ$ and $140^\circ<\theta< 165^\circ$.
Based on the QCD sum-rule technique, the decay constants $\tilde
f_s$ and $\tilde f_n$ defined by $\la f_0^q|\bar
qq|0\ra=m_{f_0}\tilde f_q$ with $f_0^n=\bar nn$ and $f_0^s=\bar
ss$ have been estimated in \cite{CYfK} by taking into account
their scale dependence and radiative corrections. It turns out
that $\tilde f_s(1\,{\rm GeV})\approx 0.33$ GeV \cite{CYfK}, for
example. In the two-quark scenario for $f_0(980)$, the decay
constants $\bar f_{n,s}$ are related to $\tilde f_{n,s}$ via
\cite{CYfK}
\begin{eqnarray}
 \bar{f}_s=\tilde{f}_s \cos\theta ,
\qquad  \bar{f}_n=\tilde{f}_n \sin\theta.
\end{eqnarray}

The hard spectator function relevant for $B\to f_0K$ decay has the
form
 \be
 H_i(f_0K) &=& {\bar f_u f_B\over
F_0^{Bf_0^u}(0)m^2_B}\int^1_0 {d\rho\over\rho}\,
\Phi_B(\rho)\int^1_0 {d\xi\over \bar\xi} \,\Phi_K(\xi)\int^1_0
{d\eta\over \deta}\left[\Phi_{f_0}(\eta)+{2m_{f_0}\over
m_b}\,{\bar\xi\over \xi}\,\Phi_{f_0}^p(\eta)\right],
 \en
for $i=1,4,10$ and $H_i=0$ for $i=6,8$. where $\bar\xi\equiv
1-\xi$ and $\bar\eta=1-\eta$. As for the parameters
$a_{6,8}^{u,c}(\ov K^0f_0)$ appearing in Eq. (\ref{eq:SDamp}),
they have the same expressions as $a_{6,8}^{u,c}(f_0\ov K^0)$
except that the penguin function $\hat G_K$ (see Eq. (55) of the
second reference in \cite{BBNS}) is replaced by $\hat G_{f_0}$ and
$\Phi^p_K$ by $\Phi^p_{f_0}$. For the distribution amplitudes
$\Phi_{f_0}$, $\Phi_{f_0}^p$ and the annihilation amplitudes, see
\cite{CYfK} for details.

Although the parameters $a_i(i\neq 6,8)$ and $a_{6,8}r_\chi$ are
formally renormalization scale and $\gamma_5$ scheme independent,
in practice there exists some residual scale dependence in
$a_i(\mu)$ to finite order. To be specific, we shall evaluate the
vertex corrections to the decay amplitude at the scale $\mu=m_b$.
In contrast, as stressed in \cite{BBNS}, the hard spectator and
annihilation contributions should be evaluated at the
hard-collinear scale $\mu_h=\sqrt{\mu\Lambda_h}$ with
$\Lambda_h\approx 500 $ MeV. There is one more serious
complication about these contributions; that is, while QCD
factorization predictions are model independent in the
$m_b\to\infty$ limit, power corrections always involve troublesome
endpoint divergences. For example, the annihilation amplitude has
endpoint divergences even at twist-2 level and the hard spectator
scattering diagram at twist-3 order is power suppressed and
possesses soft and collinear divergences arising from the soft
spectator quark. Since the treatment of endpoint divergences is
model dependent, subleading power corrections generally can be
studied only in a phenomenological way. We shall follow
\cite{BBNS} to parametrize the endpoint divergence
$X_A\equiv\int^1_0 dx/(1-x)$ in the annihilation diagram as
 \be
 X_A=\ln\left({m_B\over \Lambda_h}\right)(1+\rho_A e^{i\phi_A})
 \en
with $\rho_A\leq 1$. Likewise, the endpoint divergence $X_H$ in
the hard spectator contributions can be parametrized in a similar
way.

\subsection{Consideration of mixing-induced \CP asymmetry}

Consider the mixing-induced \CP violation in the decay modes
$(\phi,\omega,\rho^0,\pi^0,\eta',\eta,f_0)K_S$ mediated by $b\to
sq\bar q$ transitions. Since a common final state is reached only
via $K^0-\ov K^0$ mixing, hence
 \be
 \lambda_f &=& {q_B\over p_B}\,{q_K\over p_K}\,{A(\ov B^0\to M\ov
 K^0)\over A(B^0\to MK^0)} \non \\
 &\approx & {V_{tb}^*V_{td}\over
 V_{tb}V^*_{td}}\,{V_{cd}^*V_{cs}\over V_{cd}V_{cs}^*}\,{A(\ov B^0\to M\ov
 K^0)\over A(B^0\to MK^0)}.
 \en
We shall use this expression for  $\lambda_f$ to compute \CP
asymmetries $S_f$ and $\A_f$. For $M=V$, $A(B^0\to V K^0)$ has the
same expression as $-A(\ov B^0\to V\ov K^0)$ with the CKM mixing
angles $\lambda_p\to\lambda_p^*$ owing to $CP|VK^0\ra=-|V\ov
K^0\ra$, while for $M=P$, $A(B^0\to P K^0)$ is obtained from
$A(\ov B^0\to P \ov K^0)$ with $\lambda_p\to\lambda_p^*$. If the
contributions from $V_{ub}V^*_{us}$ terms are neglected, then  we
will have
 \be \label{eq:Sf}
 {A(\ov B^0\to M\ov K^0)\over A(B^0\to MK^0)}\approx
 \eta_f{V_{cd}V_{cs}^*\over V_{cd}^*V_{cs}}\quad \Rightarrow
 \lambda_f\approx \eta_f{V_{td}\over V_{td}^*}=\eta_f e^{-2i\beta} \quad\Rightarrow
 -\eta_fS_f\approx \sin 2\beta.
 \en
Note that $\eta_f=1$ for $f_0K_S$ and $\eta_f=-1$ for
$(\phi,\omega,\rho^0,\eta',\eta,\pi^0)K_S$.

From Eq. (\ref{eq:SDamp}) we see that among the seven modes under
consideration, only $\omega K_S,~\rho^0K_S,~\pi^0K_S$ and
$\eta^{(')}K_S$ receive tree contributions from the tree diagram
$b\to sq\bar q$ ($q=u,d$). However, since the tree contribution is
color suppressed, the deviation of $-\eta_fS_f$ from $\sin 2\beta$
is expected to be small. Nevertheless, the large cancellation
between $a_4$ and $a_6$ penguin terms in the amplitudes of $\ov
B^0\to\omega\ov K^0$ and $\ov B^0\to\rho^0\ov K^0$ render the tree
contribution relatively significant. Hence, $\Delta S_f$ is
expected to be largest in the $\omega K_S$ and $\rho^0K_S$ modes.
Since the typical values of the effective Wilson parameters
obtained from Eq. (\ref{eq:ai}) are
 \be
 && a_2\approx 0.18-0.11i, \qquad a_3\approx -0.003+0.005i, \qquad a_5\approx
 0.008-0.006i,  \non \\
 && a_4^u\approx -0.03-0.02i, \qquad a_4^c\approx
 -0.03-0.006i, \non \\
 && a_6^u\approx -0.06-0.02i, \qquad a_6^c\approx -0.06-0.004i,
 \en
and $r_\chi^K\approx 0.57$,  it is not difficult to see from Eq.
(\ref{eq:SDamp}) that $\delta_f$ lies in the region
$0>\delta_f>-\pi/2$ for the $\omega K_S$ and $f_0 K_S$ modes,
$\pi>\delta_f>\pi/2$ for $\rho^0K_s$ and $\pi/2>\delta_f>0$ for
the remaining three ones. Therefore, based purely on SD
contributions, it is expected that $\Delta S_f>0$ for all the
modes except for $\rho^0K_S$ and that $\A_f$ is negative for
$\omega K_S,f_0K_S$ and positive for $\phi
K_S,\rho^0K_S,\pi^0K_S,\eta^{(')}K_S$.

\subsection{Numerical results}

To proceed with the numerical calculations, we shall follow
\cite{BBNS,BNb} for the choices of the relevant parameters except
for the form factors and CKM matrix elements. For form factors we
shall use those derived in the covariant light-front quark model
\cite{CCH} and assign a common value of 0.03 for the form factor
errors, e.g. $F^{B\pi}(0)=0.25\pm0.03$. For CKM matrix elements,
see the unitarity triangle analysis in \cite{CKMfitter}. For
definiteness, we shall follow the first reference in
\cite{CKMfitter} to use the Wolfenstein parameters $A=0.801$,
$\lambda=0.2265$, $\bar\rho=0.189$ and $\bar\eta=0.358$ which
correspond to $\sin 2\beta=0.723$ and $\gamma=63^\circ$. We assign
$15^\circ$ error for the unitarity angle $\gamma$, recalling that
two values $\gamma=(62^{+10}_{-12})^\circ$ and $\gamma=(64\pm
18)^\circ$ are obtained in \cite{CKMfitter}. For endpoint
divergences encountered in hard spectator and annihilation
contributions we take the default values $\rho_A=\rho_H=0$. We
will return back to this point below when discussing long-distance
rescattering effects.

The obtained branching ratios for the decays $\ov B^0\to
(\phi,\omega,\rho^0,\pi^0,\eta',\eta,f_0)\ov K^0$ are shown in the
second column of Table \ref{tab:BR}, while the corresponding \CP
violation asymmetries $S_f$ and $\A_f$ are depicted in Table
\ref{tab:CP}. In general, our results for branching ratios and
direct \CP asymmetries are in agreement with \cite{BNb}. Some
differences result from different inputs of the form factors and
CKM parameters. It is evident that, as far as the central values
are concerned, the predicted branching ratios by the
short-distance (SD) QCD factorization approach are generally too
low compared to experiment especially for $\omega \ov K^0$,
$\rho^0K_S$ and $\pi^0\ov K^0$. Note that $\B(\ov B^0\to \omega
\ov K^0)\lsim 10^{-6}$ is predicted in the early QCD factorization
calculation \cite{Du}. The very large (small) branching ratio for
$\eta' \ov K^0$ ($\eta\ov K^0$) is understandable as follows.
There are two distinct penguin contributions to $\eta^{(')}\ov
K^0$: one couples to the $d$ quark content of the $\eta^{(')}$,
while the other is related to the $s$ quark component of the
$\eta^{(')}$ [see also Eq. (\ref{eq:SDamp})]. If the $\eta-\eta'$
mixing angle is given by $-19.5^\circ$, the expressions of the
$\eta'$ and $\eta$ wave functions will become very simple:
 \be \label{eq:etaWF}
|\eta'\ra={1\over\sqrt{6}}|\bar uu+\bar dd+2\bar ss\ra, \qquad
|\eta\ra={1\over\sqrt{3}}|\bar uu+\bar dd-\bar ss\ra.
 \en
It is evident that the SD $\eta \ov K^0$ amplitude vanishes in
SU(3) limit, whereas the constructive interference between the
penguin amplitudes accounts for the large rate of $\eta'\ov K^0$.
In reality the $\eta-\eta'$ mixing angle is $-(15.4\pm1.0)^\circ$
\cite{Feldmann}, but this does not affect the above physical
picture.

Owing to the large cancellation between the $a_4$ and $a_6$
penguin terms, the main contribution to the decay $\ov B^0\to
f_0\ov K^0$ arises from the penguin diagram involving the strange
quark content of $f_0(980)$, namely, the term with the scalar
decay constant $\bar f_s$. Consequently, the maximal branching
ratio $9.9\times 10^{-6}$ occurs near the zero mixing angle. The
result of $\B(\ov B^0\to f_0\ov K^0)=8.1\times 10^{-6}$ shown in
Table \ref{tab:BR} corresponds to $\theta=150^\circ$. Note that
the decay $\ov B^0\to f_0(980)\ov K^0$ was measured by BaBar
\cite{BaBarfK} with the result
 \be
 \B(B^0\to f_0(980)K^0\to \pi^+\pi^-K^0)=(6.0\pm0.9\pm1.3)\times
 10^{-6}.
 \en
The absolute branching ratios for $B\to f_0K$ depends critically
on the branching fraction of $f_0(980)\to \pi\pi$. We use the
results from the most recent analysis of \cite{Anisovich} to
obtain $\B(f_0(980)\to\pi\pi)=0.80\pm0.14$ and $\B(B^0\to
f_0(980)K^0)=(11.3\pm3.6)\times 10^{-6}$ as shown in Table
\ref{tab:BR}.\footnote{For comparison, the world average of the
branching ratio for $B^-\to f_0K^-\to \pi^+\pi^-K^-$ is
$(8.49^{+1.35}_{-1.26})\times 10^{-6}$ \cite{HFAG} and hence
$\B(B^-\to f_0K^-)\approx (15.9^{+3.8}_{-3.7})\times 10^{-6}$.}
In short, although the predicted branching ratios of
$(\phi,\omega,\rho^0,\eta',\pi^0)\ov K^0$ are consistent with the
data within the theoretical and experimental uncertainties, there
are sizable discrepancies between the SD theory and experiment for
the central values of their branching ratios. This may call for
the consideration of subleading power corrections such as the
annihilation topology and/or FSI effects.

In Tables \ref{tab:BR} and \ref{tab:CP} we have included the SD
theoretical uncertainties arising from the variation of the
unitarity angle $\gamma=(63\pm15)^\circ$, the renormalization
scale $\mu$ from $2m_b$ to $m_b/2$, quark masses (especially the
strange quark mass which is taken to be $m_s(2\,{\rm
GeV})=90\pm20$ MeV) and form factors as mentioned before. To
obtain the SD errors shown in Tables II and III, we first scan
randomly the points in the allowed ranges of the above four
parameters in two separated groups: the first one and the last
three ones, and then add each error in quadrature. For example,
for the decay $\ov B^0\to\eta' K_S$ we obtain
$2\B=(42.1^{+0.2+45.6}_{-0.2-19.4})\times 10^{-6}$,
$\A=1.77^{+0.30+0.22}_{-0.18-0.30}\%$ and
$S=0.737^{+0.002+0.002}_{-0.038-0.004}$, where the first error is
due to the variation of $\gamma$ and the second error comes from
the uncertainties in the renormalization scale, the strange quark
mass and the form factors.

From Table \ref{tab:CP} we obtain the differences between the \CP
asymmetry $S^{SD}_f$ induced at short distances and the measured
$S_{J/\psi K_S}$ to be
 \be \label{eq:Ssd}
 \Delta S^{SD}_{\phi K_S}=0.02^{+0.00}_{-0.04}, \qquad &&
 \Delta S^{SD}_{\omega K_S}=0.12^{+0.05}_{-0.06}, \qquad
 \Delta S^{SD}_{\rho^0K_S}=-0.09^{+0.03}_{-0.07}, \non \\
 \Delta S^{SD}_{\pi^0 K_S}=0.06^{+0.02}_{-0.04}, \qquad &&
 \Delta S^{SD}_{\eta' K_S}=0.01^{+0.00}_{-0.04}, \qquad
 \Delta S^{SD}_{\eta K_S}=0.07^{+0.02}_{-0.04}, \non \\
 && \Delta S^{SD}_{f_0K_S}=0.02^{+0.00}_{-0.04},
 \en
where the experimental error of $S_{J/\psi K_S}$ is not included.
Our results for $\Delta S^{SD}_{\phi K_S}$ and $\Delta
S^{SD}_{\eta' K_S}$ are consistent with that obtained in
\cite{BNb}.\footnote{Note that unlike \cite{BNb} we did not
include the theoretical uncertainties arising from power
corrections. Otherwise, there will be a double counting problem
when considering LD rescattering effects.}
As expected before, the $\omega K_S$ and $\rho^0K_S$ modes have
the largest deviation of $S_f$ from the naive expectation owing to
the large tree pollution. In contrast, tree pollution in $\eta'
K_S$ is diluted by the prominent $s\bar s$ content of the $\eta'$.
As for direct \CP violation, sizable direct \CP asymmetries are
predicted for $\omega K_S$ and $\rho^0K_S$ based on SD
contributions.

\begin{table}[t]
\caption{Short-distance (SD) and long-distance (LD) contributions
to the branching ratios (in units of $10^{-6}$) for various
penguin-dominated modes. The first and second theoretical errors
correspond to the SD and LD ones, respectively (see the text for
details). The world averages of experimental measurements are
taken from \cite{HFAG}.} \label{tab:BR}
\begin{ruledtabular}
\begin{tabular}{l c r r}
 & SD & SD+LD & Expt  \\ \hline
 $\ov B^0\to\phi \ov K^0$ & $5.6^{+1.9}_{-1.8}$ & $8.6^{+1.2+2.9}_{-1.2-1.8}$ & $8.3^{+1.2}_{-1.0}$ \\
 $\ov B^0\to\omega \ov K^0$ & $2.0^{+3.5}_{-1.3}$ & $5.6^{+2.9+3.7}_{-1.2-2.1}$ & $5.6\pm0.9$ \\
 $\ov B^0\to\rho^0 \ov K^0$ & $2.8^{+3.2}_{-1.6}$ &
 $5.2^{+3.2+2.6}_{-1.5-1.2}$ & $5.1\pm1.6$ \\
 $\ov B^0\to\eta' \ov K^0$ & $42.1^{+45.6}_{-19.4}$ & $69.4^{+51.3+50.4}_{-21.4-19.2}$ & $68.6\pm4.2$ \\
 $\ov B^0\to\eta \ov K^0$ & $1.8^{+1.2}_{-0.9}$ & $1.8^{+1.2+0.1}_{-0.8-0.0}$ & $<2.0$
 \\
 $\ov B^0\to\pi^0\ov K^0$ & $5.8^{+5.5}_{-3.1}$ & $9.6^{+5.5+8.4}_{-2.9-3.0}$ & $11.5\pm1.0$ \\
 $\ov B^0\to f_0\ov K^0$ & $8.1^{+3.1}_{-2.6}$ & $8.1^{+3.1+0.0}_{-2.7-0.0}$ \footnotemark[1] & $11.3\pm3.6$ \\
\end{tabular}
\end{ruledtabular}
\footnotetext[1]{Only the intermediate states $K^-\rho^+$ and
$\pi^+ K^{*-}$ are taken into account; see Sec. III for details.}
\end{table}

\begin{table}[t]
\caption{Short-distance (SD) and long-distance (LD) contributions
to the time-dependent \CP asymmetry. The first and second
theoretical errors correspond to the SD and LD ones, respectively
(see the text for details).} \label{tab:CP}
\begin{ruledtabular}
\begin{tabular}{l c c c r c c}
 &  \multicolumn{3}{c}{$-n_fS_f$}
 &   \multicolumn{3}{c}{$\A_f(\%)$}  \\ \cline{2-4} \cline{5-7}
\raisebox{2.0ex}[0cm][0cm]{Final State} & SD & SD+LD & Expt & SD &
SD+LD & Expt \\ \hline
 $\phi K_S$ & $0.747^{+0.002}_{-0.039}$ & $0.759^{+0.007+0.005}_{-0.041-0.006}$ & $0.35\pm0.20$
 & $1.4^{+0.3}_{-0.5}$ & $-2.6^{+0.8+1.1}_{-1.0-0.9}$ & $4\pm17$ \\
 $\omega K_S$ & $0.850^{+0.052}_{-0.055}$ & $0.736^{+0.022+0.025}_{-0.035-0.014}$
 & $0.55^{+0.30}_{-0.32}$ & $-7.3^{+3.5}_{-2.6}$  & $-13.2^{+3.9+2.1}_{-2.8-2.6}$ & $48\pm25$ \\
 $\rho^0K_S$ & $0.635^{+0.028}_{-0.067}$ &
 $0.761^{+0.071+0.073}_{-0.079-0.100}$ & -- & $9.0^{+2.2}_{-4.6}$ &
 $46.6^{+12.9+10.8}_{-14.7-5.9}$ & -- \\
 $\eta' K_S$ & $0.737^{+0.002}_{-0.038}$ & $0.725^{+0.004+0.005}_{-0.036-0.003}$ & $0.43\pm0.11$
 & $1.8^{+0.4}_{-0.4}$ & $2.1^{+0.6+0.5}_{-0.3-0.2}$ & $4\pm8$ \\
 $\eta K_S$ & $0.793^{+0.017}_{-0.044}$ & $0.802^{+0.025+0.002}_{-0.046-0.004}$ & $-$
 & $-6.1^{+5.1}_{-2.0}$ & $-3.7^{+4.4+1.4}_{-1.8-2.4}$ & $-$ \\
 $\pi^0K_S$ & $0.787^{+0.018}_{-0.044}$ & $0.770^{+0.006+0.015}_{-0.042-0.019}$ & $0.34^{+0.27}_{-0.29}$
 & $-3.4^{+2.1}_{-1.1}$ & $3.7^{+1.8+2.0}_{-2.0-0.4}$ & $-8\pm14$  \\
 $f_0K_S$ & $0.749^{+0.002}_{-0.039}$ & $0.749^{+0.002+0.0}_{-0.039-0.0}$ \footnotemark[1]
 & $0.39\pm0.26$ & $0.77^{+0.13}_{-0.10}$
 & $0.75^{+0.14+0.01}_{-0.09-0.01}$ \footnotemark[1] & $-14\pm22$  \\
\end{tabular}
\end{ruledtabular}
\footnotetext[1]{Only the intermediate states $K^-\rho^+$ and
$\pi^+ K^{*-}$ are taken into account (see Sec. III for details).
This means that the prediction of the LD effects on the $f_0K_S$
mode is less certain.}
\end{table}

\section{Long-distance contributions}
As noticed in passing, the predicted branching ratios for the
decays $\ov B^0\to (\phi,\omega,\rho^0,\pi^0)K_S$ by the
short-distance QCD factorization approach are generally too low by
a factor of 2 compared to experiment. Just like the pQCD approach
\cite{Keum} where the annihilation topology plays an essential
role for producing sizable strong phases and for explaining the
penguin-dominated $VP$ modes, it has been suggested in \cite{BNb}
that a favorable scenario (denoted as S4) for accommodating the
observed penguin-dominated $B\to PV$ decays and the measured sign
of direct \CP asymmetry in $\ov B^0\to K^-\pi^+$ is to have a
large annihilation contribution by choosing $\rho_A=1$,
$\phi_A=-55^\circ$ for $PP$, $\phi_A=-20^\circ$ for $PV$ and
$\phi_A=-70^\circ$ for $VP$ modes.  The sign of $\phi_A$ is chosen
so that the direct \CP violation $A_{K^-\pi^+}$ agrees with the
data. However, there are at least three difficulties with this
scenario. First, the origin of these phases is unknown and their
signs are not predicted. Second, since both annihilation and hard
spectator scattering encounter endpoint divergences, there is no
reason that soft gluon effects will only modify $\rho_A$ but not
$\rho_H$. Third, the annihilation topologies do not help enhance
the $\pi^0\pi^0$ and $\rho^0\pi^0$ modes; both pQCD and QCDF
approaches fail to describe these two color-suppressed
tree-dominated modes. As stressed in \cite{BNb}, one would wish to
have an explanation of the data without invoking weak
annihilation.

As shown in \cite{CCS}, final-state rescattering can have
significant effects on decay rates and \CP violation. For example,
the branching ratios of the penguin-dominated decay $\phi K^*$ can
be enhanced from $\sim 5\times 10^{-6}$ predicted by QCDF to the
level of $1\times 10^{-5}$ by FSIs via rescattering of charm
intermediate states \cite{CCS}. Indeed, it has been long advocated
that charming-penguin long-distance contributions increase
significantly the $B\to K\pi$ rates and yield better agreement
with experiment \cite{charmpenguin1,charmpenguin}. The
color-suppressed modes $D^0\pi^0$, $\pi^0\pi^0$ and $\rho^0\pi^0$
in $B$ decays can also be easily enhanced by rescattering effects.
Moreover, large nonperturbative strong phases can be generated
from the final-state interactions through the absorptive part of
rescattering amplitudes. We have shown explicitly in \cite{CCS}
that direct $CP$-violating partial rate asymmetries in $K^-\pi^+$,
$\rho^+\pi^-$ and $\pi^+\pi^-$ modes are significantly affected by
final-state rescattering and their signs, which are different from
what is expected from the short-distance QCDF approach, are
correctly predicted. In order to avoid the double-counting
problem, we will turn off the LD effects induced from the power
corrections due to non-vanishing $\rho_A$ and $\rho_H$; that is,
we set $\rho_A=\rho_H=0$ and $\phi_A=\phi_H=0$, therefore it is
important to note that we are not adding FSI on top of QCDF.  We
wish to stress that in principle LD rescattering effects can be
included in the framework of QCDF, but that requires modelling of
$\Lambda_{\rm QCD}/m_b$ power corrections and, in particular, one
may then need to adopt non-vanishing values of $\rho_A$, $\rho_H$,
$\phi_A$ and $\phi_H$ \cite{BNb}, as mentioned above.  In this
work, we are providing a specific model for final-state
rescattering to complement QCDF.

Besides direct \CP violation, the mixing-induced \CP asymmetry
$S_f$ also could be affected by final-state rescattering from some
intermediate states. When the intermediate states  are charmless,
the relevant CKM matrix element is $V_{ub}V_{us}^*\approx
A\lambda^4 R_be^{-i\gamma}$ which carries the weak phase $\gamma$.
In general, the charmless intermediate states will essentially not
affect the decay rates but may have potentially sizable effect on
$S_f$, whereas the charm intermediate states will affect both the
branching ratios and $S_f$.

\subsection{Final-state rescattering}

At the quark level, final-state rescattering can occur through
quark exchange and quark annihilation. In practice, it is
extremely difficult to reliably calculate the FSI effects, but it
may become amenable to estimate these effects at the hadron level
where FSIs manifest as the rescattering processes with $s$-channel
resonances and one particle exchange in the $t$-channel. In
contrast to $D$ decays, the $s$-channel resonant FSIs in $B$
decays is expected to be suppressed relative to the rescattering
effect arising from quark exchange owing to the lack of the
existence of resonances at energies close to the $B$ meson mass.
Therefore, we will model FSIs as rescattering processes of some
intermediate two-body states with one particle exchange in the
$t$-channel and compute the absorptive part of the rescattering
amplitude via the optical theorem \cite{CCS}.

Given the weak Hamiltonian in the form $H_{\rm W}=\sum_i\lambda_i
Q_i$, where $\lambda_i$ is the combination of the quark mixing
matrix elements and $Q_i$ is a $T$-even local operator ($T$: time
reversal), the absorptive part of final-state rescattering can be
obtained by using the optical theorem and time-reversal invariant
weak decay operator $Q_i$. From the time reversal invariance of
$Q~(=U_T Q^* U^\dagger_T$), it follows that
\begin{equation}
\langle i;{\rm out}| Q|B;{\rm in}\rangle^* =\sum_j S^*_{ji}
\langle j;{\rm out}|Q|B;{\rm in}\rangle,
\label{eq:timerev}
\end{equation}
where $S_{ij}\equiv\langle i;{\rm out}|\,j;{\rm in}\rangle$ is the
strong interaction $S$-matrix element, and we have used $U_T|{\rm
out\,(in)}\rangle^*=|{\rm in\,(out)}\rangle$ to fix the phase
convention.
Eq. (\ref{eq:timerev}) implies an identity related to the optical
theorem. Noting that $S=1+iT$, we find
\begin{equation}
2\,\A bs\, \langle i;{\rm out}|Q|B;{\rm in}\rangle =\sum_j
T^*_{ji} \langle j;{\rm out}|Q|B;{\rm in}\rangle, \label{eq:ImA}
\end{equation}
where use of the unitarity of the $S$-matrix has been made.
Specifically, for two-body $B$ decays, we have
 \be
 \A bs\, M(p_B\to p_ap_b) &=& {1\over 2}\sum_j\left(\Pi_{k=1}^j\int {d^3\vec
 q_k\over (2\pi)^32E_k}\right)(2\pi)^4 \non \\
 &\times & \delta^4(p_a+p_b-\sum_{k=1}^j q_k)M(p_B\to
 \{q_k\})T^*(p_ap_b\to\{q_k\}).
 \en
Thus the optical theorem relates the absorptive part of the
two-body decay amplitude to the sum over all possible $B$ decay
final states $\{q_k\}$, followed by the strong rescattering
$\{q_k\}\to p_ap_b$. In principle, the dispersive part of the
rescattering amplitude can be obtained from the absorptive part
via the dispersion relation
 \be \label{eq:dispersive}
 {\cal D}is\,A(m_B^2)={{\cal P}\over \pi}\int_s^\infty { {\A bs}\,A(s')\over
 s'-m_B^2}ds'.
 \en
Unlike the absorptive part, it is known that the dispersive
contribution suffers from the large uncertainties due to some
possible subtractions and the complication from integrations. For
this reason, we will assume the dominance of the rescattering
amplitude by the absorptive part and ignore the dispersive part in
the present work.

The relevant Lagrangian for final state strong interactions is
given by
 \be
 \L=\L_{l}+\L_h,
 \en
where
 \be
 \L_{l}&=&-\frac{1}{4}{\rm Tr}[F_{\mu\nu}(V) F^{\mu\nu}(V)]+i g_{VPP}
 {\rm Tr}(V^\mu P \lrpartial{}_{\!\mu} P)+g_{VVP}
 \epsilon^{\mu\nu\alpha\beta}{\rm Tr}(\partial_\mu V_\nu
 \partial_\alpha V_\beta P)
 \non\\
 &=&-\frac{1}{4}{\rm Tr}[f_{\mu\nu}(V) f^{\mu\nu}(V)]
 -i\frac{g_V}{2}(\phi_{\mu\nu} K^{*-\mu} K^{*+\nu}+K^{*-}_{\mu\nu}
 K^{*+\mu}\phi^\nu+K^{*+}_{\mu\nu} K^{*-\nu}\phi^\mu+\dots)
\non\\
  &&+i g_{VPP} \Big[\phi^\mu K^- \lrpartial{}_{\!\mu} K^+
               +\rho^{+\,\mu} (K^0 \lrpartial{}_{\!\mu} K^-
                               -\sqrt2 \pi^0 \lrpartial{}_{\!\mu} \pi^-)
               +K^{*-\,\mu} \pi^+ \lrpartial{}_{\!\mu} K^0
 \non\\
 &&+ K^{*-}({1\over\sqrt{2}}\eta'_q \lrpartial{}_{\!\mu} K^+ +K^+ \lrpartial{}_{\!\mu}
 \eta'_s)+ \ov K^{*0}({1\over\sqrt{2}}\eta'_q \lrpartial{}_{\!\mu} K^0+ K^0 \lrpartial{}_{\!\mu}
 \eta'_s)+\dots \Big]  \non \\
 &&+g_{VVP}~\epsilon^{\mu\nu\alpha\beta}
               \bigg[K^-(\partial_\mu K^{*+}_\nu\partial_\alpha\phi_\beta)
               +K^+(\partial_\mu\phi_\nu\partial_\alpha K^{*-}_\beta)
               +K^0(\partial_\mu K^{*-}_\nu\partial_\alpha\rho^+_\beta)
 \non\\
 &&            +\sqrt2\pi^\pm(\partial_\mu
 \rho^\mp_\nu\partial_\alpha\omega_\beta)
               +\dots\bigg],
 \label{eq:LnoD}
 \en
with $P$ and $V$ being the usual pseudoscalar and vector
multiplets, respectively,
$F_{\mu\nu}=f_{\mu\nu}+ig_V[V_\mu,\,V_\nu]/2$,
$f_{\mu\nu}=\partial_\mu V_\nu-\partial_\nu V_\mu$ and
 \be
 \L_{h}=
 &-& ig_{D^*DP}(D^i\partial^\mu P_{ij}
 D_\mu^{*j\dagger}-D_\mu^{*i}\partial^\mu P_{ij}D^{j\dagger})
 -{1\over 2}g_{D^*D^*P}
 \epsilon_{\mu\nu\alpha\beta}\,D_i^{*\mu}\partial^\nu P^{ij}
 \lrpartial{}^{\!\alpha} D^{*\beta\dagger}_j \non \\
 &-& ig_{DDV} D_i^\dagger \lrpartial_{\!\mu} D^j(V^\mu)^i_j
 -2f_{D^*DV} \epsilon_{\mu\nu\alpha\beta}
 (\partial^\mu V^\nu)^i_j
 (D_i^\dagger\lrpartial{}^{\!\alpha} D^{*\beta j}-D_i^{*\beta\dagger}\lrpartial{}{\!^\alpha} D^j)
 \non\\
 &+& ig_{D^*D^*V} D^{*\nu\dagger}_i \lrpartial_{\!\mu} D^{*j}_\nu(V^\mu)^i_j
 +4if_{D^*D^*V} D^{*\dagger}_{i\mu}(\partial^\mu V^\nu-\partial^\nu
 V^\mu)^i_j D^{*j}_\nu.
 \label{eq:LD}
 \en
Only those terms relevant for later purposes are shown in $\L_{l}$
and the  convention $\epsilon^{0123}=1$ has been adopted. For the
coupling constants, we take $g_{\rho K K}\simeq
g_{\rho\pi\pi}\simeq 4.28$,\footnote{The $\rho\pi\pi$ coupling
defined here differs from that in \cite{CCS} by a factor of
$1/\sqrt{2}$.}
$g_{\phi KK}=4.48$, $\sqrt2~ g_{VVP}=16$~GeV$^{-1}$ \cite{Bramon}.
In the chiral and heavy quark limits, we have~\cite{Casalbuoni}
 \be
 g_{D^*D^*\pi}={g_{D^*D\pi}\over m_{D^*}},
 \quad g_{DDV}=g_{D^*D^*V}=\frac{\beta g_V}{\sqrt2},
 \quad
 f_{D^*DV}=\frac{f_{D^*D^*V}}{m_{D^*}}=\frac{\lambda g_V}{\sqrt2},
 \en
with $f_\pi=132$ MeV. The parameters $g_V$, $\beta$ and $\lambda$
(not to be confused with the Wolfenstein parameter $\lambda$) thus
enter into the effective chiral Lagrangian describing the
interactions of heavy mesons with low momentum light vector mesons
(see e.g. \cite{Casalbuoni}). The parameter $g_V$ respects the
relation $g_V=m_\rho/f_\pi=5.8$ \cite{Casalbuoni}. We shall follow
\cite{Isola2003} to use $\beta=0.9$ and $\lambda=0.56$~GeV$^{-1}$.
The coupling $g_{D^*D\pi}$ has been extracted by CLEO to be
$17.9\pm0.3\pm1.9$ from the measured $D^{*+}$ width \cite{CLEOg}.

\begin{figure}[t]
  \centerline{\epsfig{figure=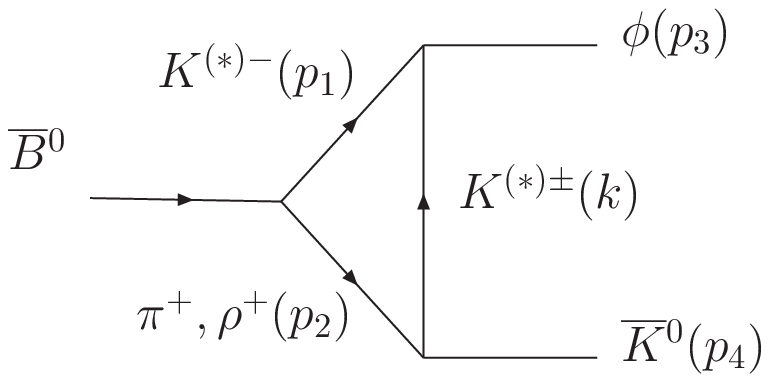,width=6cm}
              \hspace{1cm}
              \epsfig{figure=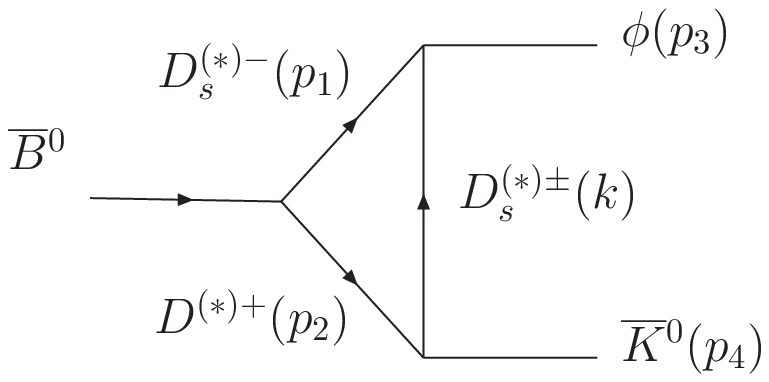,width=6cm}
              }
  \centerline{(a)
              \hspace{7cm}
              (b)
              }
    \caption{\small Final-state rescattering contributions to the
    $\overline B {}^0\to\phi \overline K {}^0$ decay.}
\end{figure}

\subsection{$B^0\to\phi K_S$ as an example}
We next proceed to study long-distance rescattering contributions
to the $b\to sq\bar q$ transition-induced decays $\ov B^0\to
(\phi,\omega,\rho^0,\eta^{(')},\pi^0,f_0)K_S$. To illustrate the
calculations of rescattering amplitudes , we shall take the $\phi
K_S$ mode as an example. Its major final-state rescattering
diagrams are depicted in Fig. 1.

The absorptive parts of the $\overline B^0\to K^-\rho^+\to\phi \ov
K {}^0$ amplitude via the $K^{\pm}$, $K^{*\pm}$ exchanges are
given by
 \be \label{eq:KrhoK}
 \A bs\,(K^-\rho^+;K^\pm) &=& {1\over 2}\int {d^3\vec p_1\over (2\pi)^32E_1}\,{d^3 \vec p_2\over
  (2\pi)^3 2E_2}\,(2\pi)^4\delta^4(p_B-p_1-p_2)
  \frac{A(\ov B^0\to K^- \rho^+)}{2\vp^*_2\cdot p_B}
\non \\
 &&\times  \sum_{\lambda_2}{2\vp^*_2\cdot p_B}\,
               (-2i) g_{\phi K K} (\vp_3^*\cdot p_1) {F^2(t,m_K)\over
 t-m_K^2}2i g_{\rho K K}\,  (\vp_2\cdot p_4)
 \non\\
 &=&2\vp^*_3\cdot p_B\times{1\over 2}\int {d^3\vec p_1\over (2\pi)^32E_1}\,{d^3 \vec p_2\over
(2\pi)^3 2E_2}\,(2\pi)^4\delta^4(p_B-p_1-p_2)
 \frac{A(\ov B^0\to K^- \rho^+)}{2\vp^*_2\cdot p_B}
\non \\
 &&\times 4g_{\phi K K}\,\frac{ F^2(t,m_K) }{t-m^2_K}\,g_{\rho K K} (A^{(1)}_1-A^{(1)}_2)
 \left(-p_1\cdot p_4+\frac{(p_1\cdot p_2)( p_2\cdot p_4)}{m_2^2}\right),
 \\
 \label{eq:KrhoKst}
  \A bs\,(K^-\rho^+;K^{*\pm}) &=& {1\over 2}\int {d^3\vec p_1\over (2\pi)^32E_1}\,{d^3 \vec p_2\over
  (2\pi)^3 2E_2}\,(2\pi)^4\delta^4(p_B-p_1-p_2)
  \frac{A(\ov B^0\to K^- \rho^+)}{2\vp^*_2\cdot p_B}
\non \\
 &&\times  \sum_{\lambda_2}{2\vp^*_2\cdot p_B}\,
               (-i) g_{\phi K^* K} [p_1,{}_\mu,p_3,\vp_3^*]
               \left(-g^{\mu\nu}+\frac{k^\mu k^\nu}{m_{K^*}^2}\right)
 \non\\
 &&\times
               \frac{F^2(t,m_{K^*})}{t-m_{K^*}^2}(-i) g_{\rho K^* K}[p_4,{}_\nu,p_2,\vp_2]
                \non\\
 &=&2\vp^*_3\cdot p_B\times{1\over 2}\int {d^3\vec p_1\over (2\pi)^32E_1}\,{d^3 \vec p_2\over
(2\pi)^3 2E_2}\,(2\pi)^4\delta^4(p_B-p_1-p_2)
 \frac{A(\ov B^0\to K^- \rho^+)}{2\vp^*_2\cdot p_B}
\non \\
 &&\times g_{\phi K^* K}\frac{ F^2(t,m_{K^*}) }{t-m^2_{K^*}}\,g_{\rho K^* K} (-2A^{(2)}_1 m_3^2),
 \en
where the dependence of the polarization vector in the amplitude
$A(\ov B^0\to K^-\rho^+)$ has been extracted and $k\equiv
p_1-p_3$. In order to avoid using too many dummy indices, we have
defined $[A,B,C,D]\equiv\epsilon_{\alpha\beta\gamma\delta}A^\alpha
B^\beta C^\gamma D^\delta$,
$[A,B,C,{}_\mu]\equiv\epsilon_{\alpha\beta\gamma\mu}A^\alpha
B^\beta C^\gamma $ and so on for later convenience. Moreover, we
have applied  the identities~\cite{CCS}
  \be
 p_{1\mu}
       &\doteq& P_\mu A_1^{(1)}+q_\mu A_2^{(1)},
 \non\\
 p_{1\mu} p_{1\nu}
       &\doteq& g_{\mu\nu} A_1^{(2)}+P_\mu P_\nu A_2^{(2)}+(P_\mu
                q_\nu+ q_\mu P_\nu) A^{(2)}_3+q_\mu q_\nu A^{(2)}_4,
 \label{eq:p1}
 \en
under the the integration
 \be
 \int {d^3\vec p_1\over (2\pi)^32E_1}\,{d^3 \vec p_2\over
(2\pi)^3
2E_2}\,(2\pi)^4\delta^4(p_B-p_1-p_2)f(t)\times\{p_{1\mu},\,p_{1\mu}
p_{1\nu}\},
 \en
with $P\equiv p_3+p_4$, $q\equiv p_3-p_4$. These identities follow
from the fact that the above integration can be expressed only in
terms of the external momenta $p_3, p_4$ with suitable Lorentz and
permutation structures. The explicit expressions of
$A^{(i)}_j=A^{(i)}_j(t,m_B^2,m_1^2,m_2^2,m_3^2,m_4^2)$ can be
found in \cite{CCS}.

Before proceeding it should be stressed that we have applied the
hidden gauge symmetry Lagrangian Eq. (\ref{eq:LnoD}) for light
vector mesons and the chiral Lagrangian Eq. (\ref{eq:LD}), based
on heavy quark effective theory (HQET) and chiral symmetry, for
heavy mesons to determine the strong vertices in Fig. 1. This
requires that the involved light pseudoscalar or vector mesons be
soft. However, the final state particles are necessarily hard and
the particle exchanged in the $t$ channel can be far off shell,
especially for the $t$-exchanged $D$ meson. This is beyond the
applicability of the aforementioned chiral perturbation theory and
HQET. Therefore, as stressed in \cite{CCS}, it is necessary to
introduce the form factor $F(t,m)$ appearing in Eqs.
(\ref{eq:KrhoK}) and (\ref{eq:KrhoKst}) to take care of the
off-shell effect of the $t$-channel exchanged particle and the
hardness of the final particles. Indeed, if the off-shell effect
is not considered, the long-distance rescattering contributions
will become so large that perturbation theory is no longer
trustworthy. For example, since $B\to D_s\bar D$ is CKM doubly
enhanced relative to $B\to K\pi$, the rescattering process $B\to
D_s\bar D\to K\pi$ will overwhelm the initial $B\to K\pi$
amplitude. Hence, form factors or cutoffs must be introduced to
the strong vertices to render the calculation meaningful in
perturbation theory.

The form factor $F(t,m)$ is usually parametrized as
 \be \label{eq:FF}
 F(t,m)=\,\left({\Lambda^2-m^2\over \Lambda^2-t}\right)^n,
 \en
normalized to unity at $t=m^2$ with $\Lambda$ being a cutoff
parameter which should be not far from the physical mass of the
exchanged particle. To be specific, we write
 \be \label{eq:Lambda}
 \Lambda=m_{\rm exc}+\eta\Lambda_{\rm QCD},
 \en
where the parameter $\eta$ is expected to be of order unity and it
depends not only on the exchanged particle but also on the
external particles involved in the strong-interaction vertex. The
monopole behavior of the form factor (i.e. $n=1$) is preferred as
it is consistent with the QCD sum rule expectation
\cite{Gortchakov}. Although the strong couplings are large in the
magnitude, the rescattering amplitude is suppressed by a factor of
$F^2(t)\sim m^2\Lambda_{\rm QCD}^2/t^2$. Consequently, the
off-shell effect will render the perturbative calculation
meaningful. Moreover, since in the heavy quark limit $t\sim
m_B^2$, the final-state rescattering amplitude does vanish in the
$m_B\to\infty$ limit, as it should.

Likewise, the absorptive part of the $\overline B^0\to
K^{*-}\pi^+\to\phi \ov K {}^0$ amplitude via the $K^{*\pm}$
exchange is given by
 \be \label{eq:KstpiKst}
 \A bs\,(K^{*-}\pi^+;K^{*\pm}) &=& {1\over 2}\int {d^3\vec p_1\over (2\pi)^32E_1}\,{d^3 \vec p_2\over
  (2\pi)^3 2E_2}\,(2\pi)^4\delta^4(p_B-p_1-p_2)
  \frac{A(\ov B^0\to K^{*-} \pi^+)}{2\vp^*_1\cdot p_B}
\non \\
 &&\times  \sum_{\lambda_1}{2\vp^*_1\cdot p_B}\,
               i \frac{g_V}{2}[\vp_{3\mu}^*(2 \vp_1\cdot p_3)
               -\vp_1\cdot \vp_3^*(p_1+p_3)_\mu+\vp_{1\mu}(2 p_1\cdot\vp_3)] \non\\
 &&\times
               \left(-g^{\mu\nu}+\frac{k^\mu k^\nu}{m_{K^*}^2}\right)
               \frac{F^2(t,m_{K^*})}{t-m_{K^*}^2}
               (-i) g_{K^* K\pi}(p_2+p_4)_\nu \\
 &=&2\vp^*_3\cdot p_B\times{1\over 2}\int {d^3\vec p_1\over (2\pi)^32E_1}\,{d^3 \vec p_2\over
(2\pi)^3 2E_2}\,(2\pi)^4\delta^4(p_B-p_1-p_2)
 \frac{A(\ov B^0\to K^{*-} \pi^+)}{2\vp^*_1\cdot p_B}
\non \\
 &&\times ~g_V \frac{F^2(t,m_{K^*}) }{2(t-m_{K^*}^2)}\,g_{K^* K\pi}
               \Bigg\{2\left[2-A^{(1)}_1+A^{(1)}_2+(A^{(1)}_1-A^{(1)}_2)\frac{m_2^2-m_4^2}{m_{K^*}^2}\right]
               \non\\
 &&\times  \left(p_2\cdot p_3-\frac{(p_1\cdot p_2)(p_2\cdot p_3)}{m_1^2}\right)
           +\left[A^{(1)}_1-A^{(1)}_2-1+(A^{(1)}_1-A^{(1)}_2)
           \frac{p_1\cdot p_2}{m_1^2}\right]
           \non\\
 &&\times  \left[(p_1+p_3)\cdot (p_2+p_4)+\frac{1}{m_{K^*}^2}(m_1^2-m_3^2)(m_2^2-m_4^2)\right]
           +2(A^{(1)}_1-A^{(1)}_2)
           \non\\
 &&\times  \left[\left(p_2\frac{m_{K^*}^2-m_2^2+m_4^2}{m_{K^*}^2}+p_4\frac{m_{K^*}^2+m_2^2-m_4^2}{m_{K^*}^2}\right)
 \cdot\left(p_2-p_1 \frac{p_1\cdot p_2}{m_1^2}\right)
           \right]\Bigg\}. \non
 \en
Note that since the $\pi K K$ vertex is absent, there is no
contribution from the $K^+$ exchanged particles.

The $\ov B^0\to \phi \ov K^0$ decay also receives contributions
from charmless $VV$ modes. The leading candidate is the
$K^{*-}\rho^+$ mode via the $p$-wave configuration. However, as we
have checked numerically, its amplitude is one or two orders of
magnitude smaller than those from the previous two charmless $VP$
modes and its effect on $S_f$ is quite small.  Given this, we will
not go into any further detail on the rescattering from charmless
$VV$ modes. Thus far we have only considered the contributions
where the two intermediate states originating from the weak vertex
are on shell. There are additional contributions where one of the
mesons coming from the weak vertex and the exchanged meson in the
$t$-channel are on shell. For example, in the diagram of Fig.
1(a), we can set $K^-$ and $K^+$ on shell while keeping $\rho^+$
off shell. This corresponds to the 3-body weak decay $\ov B^0\to
K^+K^-\ov K^0$ followed by the strong rescattering $K^+K^-\ov
K^0\to \phi \ov K^0$ where $\ov K^0$ behaves as a spectator.
However, there are many possible pole contributions to $\ov B^0\to
K^+K^-\ov K^0$. In addition to $\ov B^0\to K^-\rho^+\to K^+K^-\ov
K^0$ as inferred from Fig. 1(a), one can also have $\ov B^0\to \ov
B_s^*\ov K^0\to K^+K^-\ov K^0$, for example. In the present work,
we will only focus on two-body intermediate state contributions to
the absorptive part. Since the analogous 3-body contributions do
not occur in Fig. 1(b) and since Fig. 1(a) is CKM doubly
suppressed relative to Fig. 1(b), it is safe to neglect the
additional three-body contributions for our purposes.

We next turn to the FSI contribution arising from the intermediate
states $D_s^{(*)-} D^{(*)}$. Note that only the $p$-wave
configurations of the $D_s^{(*)-} D^{(*)}$ systems can rescatter
into the $\phi \ov K {}^0$ final state.
The absorptive parts of $\overline B {}^0\to D^{*-}_s D^+\to\phi
\overline K {}^0$ amplitudes via $D^*_s$ exchanges, $\overline
B\to D^{*-}_s D^+,\,D^{*-}_s D^{*+}\to\phi \overline K {}^0$
amplitudes via $D_s$ and $D^*_s$ exchanges, are given by
 \be
  \A bs\,(D^{*-}_s D^+;D^{*\pm}_s) &=&
                    {1\over 2}\int {d^3\vec p_1\over (2\pi)^32E_1}\,{d^3 \vec p_2\over (2\pi)^3 2E_2}\,
                    (2\pi)^4\delta^4(p_B-p_1-p_2)\frac{A(\ov B^0\to D^{*-}_s D^+)}{2\vp^*_1\cdot P}
\non \\
 &&\times  \sum_{\lambda_1}{2\vp^*_1\cdot P}\,(-4i) f_{D^*_s D^*_s \phi}
 {F^2(t,m_{D^*_s})\over t-m_{D^*_s}^2}i\,g_{D_s^* D K}\,
 \non\\
 &&\times\vp_1^\rho(\sigma g_{\rho\mu} p_1\cdot\vp^*_3+p_{3\rho}\vp_{3\mu}^*-\vp_{3\rho}^*p_{3\mu})
          \bigg(-g^{\mu\nu}+\frac{k^\mu k^\nu}{m_{D^*_s}^2}\bigg) p_{4\nu}
 \non\\
 &=&2 \vp_3^*\cdot p_B\times{1\over 2}\int {d^3\vec p_1\over (2\pi)^32E_1}\,{d^3 \vec p_2\over
   (2\pi)^3 2E_2}\,(2\pi)^4\delta^4(p_B-p_1-p_2)
\non \\
 &&\times\frac{A(\ov B^0\to D^{*-}_s D^+)}{2\vp^*_1\cdot P}
         ~4\,f_{D^*_s D^*_s \phi} {F^2(t,m_{D^*_s})\over t-m_{D^*_s}^2} g_{D_s^* D K}
 \non\\
 &&\times\bigg\{\sigma (A^{(1)}_1-A^{(1)}_2)
  \bigg(p_2-\frac{p_1\cdot p_2}{m_1^2} p_1\bigg)\cdot
          \bigg(p_4-k\frac{k\cdot p_4}{m_{D_s^*}^2}\bigg)
 \non\\
 &&       +\bigg[1-(A^{(1)}_1-A^{(1)}_2)\frac{k\cdot p_4}{m_{D_s^*}^2}\bigg]
           \bigg(p_2\cdot p_3-\frac{p_1\cdot p_2~p_1\cdot  p_3}{m_1^2}\bigg)
 \non\\
 &&         -\bigg[1-(A^{(1)}_1-A^{(1)}_2)\frac{P\cdot p_1}{m_1^2}\bigg]
           \bigg(p_3\cdot p_4-\frac{p_3\cdot k~k\cdot p_4}{m_{D_s^*}^2}\bigg)\bigg\},
 \non\\
 %
 %
 %
 %
 \A bs\,(D^-_s D^{*+};D^{\pm}_s)
  &=&{1\over 2}\int {d^3\vec p_1\over (2\pi)^32E_1}\,{d^3 \vec p_2\over (2\pi)^3 2E_2}\,
     (2\pi)^4\delta^4(p_B-p_1-p_2)\frac{A(\ov B^0\to D^-_s D^{*+})}{2\vp^*_2\cdot P}
 \non \\
 &&\times  \sum_{\lambda_2}{2\vp^*_2\cdot P}\,
           2i g_{D_s D_s \phi}\, \vp_3^*\cdot p_1\, {F^2(t,m_{D_s})\over t-m_{D_s}^2}\,(-i) g_{D^* D_s K}\,p_4\cdot \vp_2
 \non\\
 &=&2\vp_3^*\cdot p_B\times{1\over 2}\int {d^3\vec p_1\over (2\pi)^32E_1}\,{d^3 \vec p_2\over
    (2\pi)^3 2E_2}\,(2\pi)^4\delta^4(p_B-p_1-p_2)\frac{A(\ov B^0\to D^-_s D^{*+})}{2\vp^*_2\cdot P}
\non \\
 &&\times \bigg\{2 g_{D_s D_s \phi} {F^2(t,m_{D_s})\over t-m_{D_s}^2} g_{D^* D_s K}\,
           (A^{(1)}_1-A^{(1)}_2)\bigg(-p_1\cdot p_4+\frac{p_1\cdot p_2~p_2\cdot p_4}{m_2^2}\bigg)\bigg\},
 \non\\
 \A bs\,(D_s^-D^{*+};D^{*\pm}_s)
 &=& {1\over 2}\int {d^3\vec p_1\over (2\pi)^32E_1}\,{d^3 \vec p_2\over (2\pi)^3 2E_2}\,
     (2\pi)^4\delta^4(p_B-p_1-p_2)\frac{A(\ov B^0\to D^-_s D^{*+})}{2\vp^*_2\cdot P}
\non \\
 &&\times  \sum_{\lambda_2}{2\vp^*_2\cdot P}\,4i\, f_{D^*_s D_s \phi} {F^2(t,m_{D^*_s})\over
 t-m_{D^*_s}^2}ig_{D^* D_s^* K}\,
 \non\\
 &&\times [p_3,\vp^*_3,p_1,{}_\mu]\bigg(-g^{\mu\nu}+\frac{k^\mu k^\nu}{m_{D^*_s}^2}\bigg)
          [\vp_2, p_4, p_2, {}_\nu]
 \non\\
 &=&2\vp_3^*\cdot p_B\times{1\over 2}\int {d^3\vec p_1\over (2\pi)^32E_1}\,{d^3 \vec p_2\over(2\pi)^3 2E_2}\,
     (2\pi)^4\delta^4(p_B-p_1-p_2)\frac{A(\ov B^0\to D^-_s D^{*+})}{2\vp^*_2\cdot P}
\non \\
 &&\times ~(-8)\,f_{D^*_s D_s \phi} {F^2(t,m_{D^*_s})\over
 t-m_{D^*_s}^2}g_{D^* D_s^* K}~m_3^2 A^{(2)}_1,
 \non\\
 \A bs\,(D^{*-}_s D^{*+};D^{\pm}_s)
 &=& {1\over 2}\int {d^3\vec p_1\over (2\pi)^32E_1}\,{d^3 \vec p_2\over (2\pi)^3 2E_2}\,
 (2\pi)^4\delta^4(p_B-p_1-p_2)(ic[{}_\mu,{}_\nu,P,p_2])
\non \\
 &&\times  \sum_{\lambda_1,\lambda_2}{\vp^{*\mu}_1\vp^{*\nu}_2}\,
           (-4i) f_{D^*_s D_s \phi} {F^2(t,m_{D_s})\over t-m_{D_s}^2}\,(-i)g_{D^* D_s K}\,
            [p_3,\vp^*_3,p_1,\vp_1] p_4\cdot \vp_2
 \non\\
 &=&2\vp_3^*\cdot p_B\times{1\over 2}\int {d^3\vec p_1\over (2\pi)^32E_1}\,{d^3 \vec p_2\over(2\pi)^3 2E_2}\,
   (2\pi)^4\delta^4(p_B-p_1-p_2)
 \non\\
   &&\times(-4ic) f_{D^*_s D_s \phi} {F^2(t,m_{D_s})\over t-m_{D_s}^2} g_{D^* D_s K}m_3^2 A^{(2)}_1,
 \non\\
  %
  %
  %
 \A bs\,(D^{*-}_s D^{*+};D^{*\pm}_s)
 &=& {1\over 2}\int {d^3\vec p_1\over (2\pi)^32E_1}\,{d^3 \vec p_2\over (2\pi)^3 2E_2}\,
        (2\pi)^4\delta^4(p_B-p_1-p_2)(ic[{}_\delta,{}_\beta,P,p_2])
 \non \\
 &&\times  \sum_{\lambda_1,\lambda_2}{\vp^{*\delta}_1\vp^{*\beta}_2}\,
           (-4i) f_{D^*_s D^*_s \phi} {F^2(t,m_{D^*_s})\over t-m_{D^*_s}^2}i g_{D^* D^*_s K}
 \non\\
 &&\times \vp_1^\rho(\sigma g_{\rho\mu}p_1\cdot\vp^*_3\!\!+p_{3\rho}\vp_{3\mu}^*\!\!-\vp_{3\rho}^*p_{3\mu})
          \bigg(-g^{\mu\nu}+\frac{k^\mu k^\nu}{m_{D^*_s}^2}\bigg)
          [\vp_2,p_4,p_2,{}_\nu]
 \non\\
 &=&2\vp^*_3\cdot p_B\times{1\over 2}\int {d^3\vec p_1\over (2\pi)^32E_1}\,{d^3 \vec p_2\over (2\pi)^3 2E_2}\,
 (2\pi)^4\delta^4(p_B-p_1-p_2)
 \non\\
 &&\times(-4i c) f_{D^*_s D^*_s \phi} {F^2(t,m_{D^*_s})\over t-m_{D^*_s}^2} g_{D^* D^*_s K^*}
\non \\
 &&\times\big[\sigma (A^{(1)}_1-A^{(1)}_2)(p_1\cdot p_4
 m_2^2-p_1\cdot p_2 p_2\cdot p_4)+m_3^2 A^{(2)}_1\big],
 \label{eq:DsD}
  \en
where the dependence of the polarization vectors in $A(\ov B^0\to
D^{*-}_s D^+,\,D^-_s D^{*+})$ has been extracted out explicitly
and $\sigma\equiv g_{D^*D^*V}/(2f_{D^*D^*V})$ and the $\overline
B\to D^{*-}_s D^{*+}$ decay amplitude has been denoted as
 \be
 A(\overline B\to D^*_s(p_1,\lambda_1) D^*(p_2,\lambda_2))=\vp^{*\mu}_1\vp^{*\nu}_2(a\,
 g_{\mu\nu}+bP_\mu P_\nu+ic\,[{}_\mu,{}_\nu,P,p_2]).
 \en
In order to perform a numerical study of  the above analytic
results, we need to specify the short-distance $A(\ov B\to
D_s^{(*)} D^{(*)})$ amplitudes. In the factorization approach, we
have
 \be
 A(\ov B\to D^*_s D)_{SD} &=&\frac{G_{\rm F}}{\sqrt2} V_{cb} V^*_{cs} a_1
 f_{D^*_s} m_{D_s^*} F_1^{BD}(m_{D^*_s}^2)(2\vp^*_{D_s^*}\cdot p_B),
 \non\\
 A(\ov B\to D_s D^*)_{SD} &=&\frac{G_{\rm F}}{\sqrt2} V_{cb} V^*_{cs} a_1
 f_{D_s} m_{D^*} A_0^{BD^*}(m_{D_s}^2)(2\vp^*_{D^*}\cdot p_B),
 \non\\
 A(\ov B\to D^*_s D^*)_{SD} &=& -i\frac{G_{\rm F}}{\sqrt2} V_{cb} V^*_{cs} a_1
 f_{D^*_s} m_{D_s^*} (m_B+m_{D^*})\vp^{*\mu}_{D_s^*}\vp^{*\nu}_{D^*}
 \bigg[A_1^{BD^*}(m_{D^*_s}^2) g_{\mu\nu}
 \non\\
 &&-\frac{2 A_2^{BD^*}(m_{D^*_s}^2)}{(m_B+m_{D^*})^2}p_{B\mu} p_{B\nu}
 -i\frac{2 V^{BD^*}(m_{D^*_s}^2)}{(m_B+m_{D^*})^2}
 \epsilon_{\mu\nu\alpha\beta}p_B^\alpha p_{D^*}^\beta \bigg].
 \en
Since the phase of the parameter $a_1$ originating from vertex
corrections [see Eq. (\ref{eq:ai})] is very small, one can neglect
the strong phase of the short-distance amplitudes.

The long-distance contribution to the $\ov B {}^0\to \omega\ov
K{}^0$ decay can be preformed similarly. Due to the absence of the
$PPP$ vertex and the $G$-parity argument, the number of FSI
diagrams from charmless intermediate states is greatly reduced
compared to the previous case. For example, the $K^-\rho^+$
intermediate state does not contribute to the $\ov K {}^0\omega$
amplitude (through $t$-channel $\pi$ and $\rho$ exchanges) as both
$K K \pi$ and $\rho \rho \omega$ vertices are forbidden.
In fact, there is only one relevant rescattering diagram, namely,
$\ov B^0\to K^{*-}\pi^+\to \ov K^0\omega$ via $\rho^\pm$ exchange,
arising from charmless intermediate states and the corresponding
absorptive part is given by
 \be \label{eq:Kstpirho}
 \A bs\,(K^{*-}\pi^+;\rho^\pm) &=& {1\over 2}\int {d^3\vec p_1\over (2\pi)^32E_1}\,{d^3 \vec p_2\over
  (2\pi)^3 2E_2}\,(2\pi)^4\delta^4(p_B-p_1-p_2)
  \frac{A(\ov B^0\to K^{*-} \pi^+)}{2\vp^*_1\cdot p_B}
\non \\
 &&\times  \sum_{\lambda_1}{2\vp^*_1\cdot p_B}\,
               (i g_{\rho K^* K}) (i{\sqrt2}g_{\omega\rho\pi})\,
               [p_1,\vp_1,k,{}_\mu][k,{}_\nu,p_4,\vp_4^*]
               \non\\
 &&\times
               \left(-g^{\mu\nu}+\frac{k^\mu k^\nu}{m_{K^*}^2}\right)
               \frac{F^2(t,m_\rho)}{t-m_\rho^2}
                \non\\
 &=&2\vp^*_4\cdot p_B\times{1\over 2}\int {d^3\vec p_1\over (2\pi)^32E_1}\,{d^3 \vec p_2\over
(2\pi)^3 2E_2}\,(2\pi)^4\delta^4(p_B-p_1-p_2)
 \frac{A(\ov B^0\to K^{*-} \pi^+)}{2\vp^*_1\cdot p_B}
\non \\
 &&\times (-2\sqrt2)\,g_{\rho K^* K}
 \frac{F^2(t,m_\rho) }{(t-m_\rho^2)}\,g_{\omega\rho\pi}
               m_4^2 A^{(2)}_1.
 \en
Furthermore, we have checked numerically that the $\ov B^0\to
K^{*-}\rho^+\to\ov K^0\omega$ contribution is small. Hence we will
skip further detail on the rescattering from charmless $VV$ modes.

The rescattering amplitudes of $\overline B {}^0\to D^{*-}_s
D^+,\,D^- D^{*+},\,D^{*-}D^{*+}\to\ov K {}^0\omega$ amplitudes via
$D,\,D^*$ exchanges can be evaluated in a similar way. The
analytic expressions of their absorptive parts are similar to
those for $\phi\ov K {}^0$. For example,
 \be
  \A bs\,(D^{*-}_s D^+;D^{\pm}) &=&
2\vp_4^*\cdot p_B\times{1\over 2}\int {d^3\vec p_1\over
(2\pi)^32E_1}\,{d^3 \vec p_2\over
    (2\pi)^3 2E_2}\,(2\pi)^4\delta^4(p_B-p_1-p_2)\frac{A(\ov B^0\to D^-_s D^{*+})}{2\vp^*_2\cdot P}
\non \\
 &&\times \bigg\{\sqrt2 g_{D_s^* D K} {F^2(t,m_{D})\over t-m_{D}^2} g_{D D \omega}\,
           (1-A^{(1)}_1-A^{(1)}_2)\bigg(-p_2\cdot p_3+\frac{p_1\cdot p_2~p_1\cdot p_3}{m_1^2}\bigg)\bigg\},
\non\\
  \en
is the same as $\A bs(D_s^- D^{*+};D_s^\pm)$ in (\ref{eq:DsD})
after the replacements $g_{D^* D_s K}\to g_{D_s^* D K}$, $f_{D_s
D_s\phi}\to f_{D D \omega}/\sqrt2$, $p_1\leftrightarrow p_2$,
$p_3\leftrightarrow p_4$, $A^{(1)}_1-A^{(1)}_2\to
1-A^{(1)}_1-A^{(1)}_2$ and a suitable replacement of the source
amplitude (note that the replacement of momentum should not be
performed in $A^{(i)}_j$). Likewise, we can obtain $\A bs(D^{*-}_s
D^+;D^{*\pm})$, $\A bs(D^{-}_s D^{*+};D^{*\pm})$, $\A bs(D^{*-}_s
D^{*+}, D^{(*)\pm})$ from $\A bs(D^{-}_s D^{*+};D^{*\pm}_s)$, $\A
bs(D^{*-}_s D^{+};D^{*\pm}_s)$, $\A bs(D^{*-}_s D^{*+},
D^{(*)\pm}_s)$ in (\ref{eq:DsD}), respectively, with $A^{(2)}_1$
being unchanged and an additional overall minus sign for $D^{*-}_s
D^{*+}$ contributions.
This similarity is by no means accidental; it follows from the
so-called CPS symmetry, i.e. CP plus $s\leftrightarrow d$ switch
symmetry. A similar but more detailed discussion is given in
\cite{CCS} for the case of $B\to \phi K^*$ decay.

The final-state rescattering contributions to other
penguin-dominated decays $B^0\to (\rho^0,\eta^{(')},f_0)K_S$ can
be worked out in a similar manner. The dominant intermediates
states for each decay modes are summarized in Table
\ref{tab:intermediatestate}.

\subsection{Results and discussions}

\begin{table}[t]
\caption{Dominant intermediate states contributing to various
final states. For the $\eta K_S$ final state, the intermediate
states are the same as that for $\eta' K_S$.}
\label{tab:intermediatestate}
\begin{center}
\begin{tabular}{|c l|}
 \hline
 ~~~Final state\qquad\qquad
      & Intermediate state (exchanged particle)
      \\
 \hline
 $\phi K_S$
      & $K^{*-}\pi^+(K^*),$
        $K^-\rho^+(K,K^*),$
      \\
      & $D_s^{*-} D^+ (D_s^*),$
        $D_s^- D^{*+} (D_s,D_s^*),$
        $D_s^{*-} D^{*+} (D_s,D_s^*)$~~~
        \\
 $\omega K_S$
      & $K^{*-}\pi^+(\rho),$
      \\
      & $D_s^{*-} D^+ (D,D^*),$
        $D_s^- D^{*+} (D^*),$
        $D_s^{*-} D^{*+} (D,D^*)$
        \\
 $\rho^0 K_S$
      & $K^{*-}\pi^+(\rho),$
      \\
      & $D_s^{*-} D^+ (D,D^*),$
        $D_s^- D^{*+} (D^*),$
        $D_s^{*-} D^{*+} (D,D^*)$
        \\
 $\eta'K_S$
      & $\eta'\bar K^0(K^{*0})$,
        $K^{*-}\rho^+(K,K^*)$
      \\
      & $D_s^-D^+(D^*, D_s^*)$,
        $D_s^{*-}D^{*+}(D,D_s,D^*,D_s^*)$
        \\
 $\pi^0K_S$
      & $\eta'\bar K^0(K^{*0})$
      \\
      & $D_s^-D^+(D^*)$,
        $D_s^{*-}D^{*+}(D,D^*)$
        \\
 $f_0K_S$
      & $\rho^+ K^-(K,\rho)$,
        $\pi^+ K^{*-}(\pi)$
      \\
      & $D_s^{*-}D^+(D,D_s^*)$,
        $D_s^{-}D^{*+}(D^*)$
        \\
        \hline
\end{tabular}
\end{center}
\end{table}

Writing $A=A^{SD}+i \a bs A^{LD}$ with $\a bs A^{LD}$ obtained
above and the form factors given in \cite{CCH}, the results of the
final-state rescattering effects on decay rates, direct and
mixing-induced \CP violation parameters are shown in Tables
\ref{tab:BR} and \ref{tab:CP}. As pointed out in \cite{CCS}, the
long-distance rescattering effects are sensitive to the cutoff
parameter $\Lambda$ appearing in Eq. (\ref{eq:Lambda}) or $\eta$
in Eq. (\ref{eq:Lambda}). Since we do not have first-principles
calculations of $\eta$, we will determine it from the measured
branching ratios and then use it to predict the $CP$-violating
parameters $\A_f$ and $S_f$. As shown in \cite{CCS}, $\eta=0.69$
for the exchanged particles $D$ and $D^*$ is obtained from fitting
to the $B\to K\pi$ rates. We take $\eta=0.85$ for the exchanged
particles $D_{(s)}$ and $D^*_{(s)}$ to fit the data of $\ov B^0\to
\eta'\ov K^0$ rates and $\eta=1$ for other light exchanged
particles such as $\pi,K,K^*$. As for the $VP$ modes, namely,
$\phi K_S$, $\omega K_S$ and $\rho^0K_S$, we take
$\eta_{D_{(s)}}=\eta_{D^*_{(s)}}=0.95,~1.4$ and 1.5 respectively
in order to accommodate their rates~\footnote{Note that a monopole
momentum dependence is used for the form factors throughout this
paper [see Eq. (\ref{eq:FF})]. If a dipole form of the momentum
dependence is used for $D^*$-exchange diagrams as in \cite{CCS},
we will obtain
 ${\mathcal B}^{SD+LD}(\phi K^0)=(6.1^{+2.0+0.4}_{-1.9-0.2})\times 10^{-6}$,
 $S^{SD+LD}_{\phi K_S}=0.723^{+0.004}_{-0.036} {}^{+0.009}_{-0.012}$,
 $\A^{SD+LD}_{\phi K_S}=-0.070^{+0.015}_{-0.017}\pm0.019$,
 ${\mathcal B}^{SD+LD}(\omega K^0)=(2.5^{+3.7}_{-1.4}{}^{+0.6}_{-0.3})\times 10^{-6}$,
 $S^{SD+LD}_{\omega K_S}=0.847^{+0.029}_{-0.054}{}^{+0.007}_{-0.013}$ and
 $\A^{SD+LD}_{\omega K_S}=0.013^{+0.034}_{-0.078}{}^{+0.027}_{-0.029}$.}.
Note that the result $\eta_{D_{(s)}}=\eta_{D^*_{(s)}}=1.5$ for
$\rho^0 K_S$ is consistent with the value of
$\eta_{D}=\eta_{D^*}=1.6$ obtained for $B\to \rho\pi$ decays
\cite{CCS} within SU(3) symmetry. For the decay $\ov B^0\to f_0
K_S$, since only the strong couplings of $f_0$ to $K\bar K$ and
$\pi\pi$ are available experimentally, we shall only consider the
intermediate states $K^-\rho^+$ and $\pi^+K^{*-}$. The LD
theoretical uncertainties shown in Tables \ref{tab:BR} and
\ref{tab:CP} originate from three sources: an assignment of 15\%
error in $\Lambda_{\rm QCD}$, the measured error in the coupling
$g_{D^*D\pi}=17.9\pm0.3\pm 1.9$ \cite{CLEOg} and 5\% error in the
form factors for $B$ to $D^{(*)}$ transitions. They are obtained
by scanning randomly the points in the allowed ranges of the
above-mentioned three parameters. The calculations of hadronic
diagrams for FSIs also involve many other theoretical
uncertainties some of which are already discussed in \cite{CCS}.
From Tables \ref{tab:BR} and \ref{tab:CP} it is clear that the SD
errors are in general not significantly affected by FSI effects
and that LD uncertainties are in general comparable to the SD
ones.

We see from Table \ref{tab:BR} that final-state rescattering will
enhance the decay rates of $\phi K_S,\omega
K_S,\rho^0K_S,\eta'K_S$ and $\pi^0K_S$ but it does not affect the
$\eta K_S$ rate. The seemingly large disparity between $\eta'K_S$
and $\eta K_S$ for FSIs can be understood as follows. There are
two types of exchanged particles in the rescattering processes,
namely, $D^{(*)}$ and $D_s^{(*)}$ (see Table
\ref{tab:intermediatestate}). The former (latter) couple to the
$d$ ($s$) quark component of the $\eta^{(')}$. Since the $\eta'$
and $\eta$ wave functions are approximately given by Eq.
(\ref{eq:etaWF}), it is clear that the rescattering amplitudes due
to the exchanged particles $D^{(*)}$ and $D_s^{(*)}$ interfere
constructively for the $\eta'K_S$ production but compensate
largely for $\eta K_S$.

It should be stressed that although we have used the measured
branching ratio of $\eta'K^0$ to fix the LD contributions and the
unknown cutoff parameter $\eta$, there exist some other possible
mechanisms that can help explain its large rate. For example, the
QCD anomaly effect manifested in the two gluon coupling with the
$\eta'$ may provide a dynamical enhancement of the $\eta' K^0$
production \cite{AS97}. And it is likely that both final-state
rescattering and the gluon anomaly are needed to account for the
unexpectedly large branching fraction of $\eta'K$. Note that both
contributions carry negligible $CP$-odd phase. Hence, whether the
anomalously large branching ratio of $\eta'K$  comes from the QCD
anomaly and/or from final-state rescattering, it will be very
effective in diluting the $u\bar u$ tree contributions and
rendering $\Delta S_{\eta'K_S}$ small.

We are not able to estimate the long-distance rescattering
contributions to the $f_0K_S$ rate from intermediate charm states
due to the absence of information on $f_0DD$ and $f_0
D^*_{(s)}D^*_{(s)}$ couplings.

Since the modes $\omega K_S,\rho^0K_S,\phi K_S$ and $\pi^0K_S$
receive significant final-state rescattering contributions, it is
natural to expect that their direct \CP asymmetries will be
affected accordingly. It is clear from Table \ref{tab:CP} that the
signs of $\A_f$ in the last two channels are flipped by
final-state interactions. As for the mixing-induced \CP violation
$S_f$, we see from Table \ref{tab:CP} that $\omega K_S$ and
$\rho^0K_S$ receive the largest corrections from final-state
rescattering, while the long-distance correction to $\phi K_S$ is
not as large as what was originally expected. The underlying
reason is as follows.
The mixing \CP asymmetries $S^{SD}_{\omega K_s}$ and
$S^{SD}_{\rho^0K_S}$ deviate from $\sin 2\beta$ as they receive
contributions from the tree amplitude. The contribution from the
tree amplitude is relatively enhanced  as the QCD penguin
amplitude is suppressed by a cancellation between penguin terms
($|a_4-r_\chi a_6|\ll |a_4|,\,|a_6|$). Final-state rescattering
form $D^{(*)}_s D^{(*)}$ states with vanishing weak phases will
dilute the $SD$ tree contribution and bring the asymmetry closer
to $\sin 2\beta$. For the $\phi K_S$ mode, the asymmetry
$S^{SD}_{\phi K_S}$ is slightly greater than $\sin 2\beta$. With
rescattering from either charmless intermediate states, such as
$K^*\pi$, or from $D^{(*)}_s D^{(*)}$ states, the asymmetry is
reduced to $S_{\phi K_S}\simeq 0.73$. When both charmless and
charmful intermediate states are considered, the asymmetry is
enhanced to $S_{\phi K_S}\simeq 0.76$ owing to the interference
effect from these two contributions.

Among the seven modes
$(\phi,\omega,\rho^0,\pi^0,\eta^{(')},f_0)K_S$, the first three
are the states where final-state interactions may have a
potentially large effect on the mixing-induced \CP asymmetry
$S_f$. In order to maximize the effects of FSIs on $S_f$, one
should consider rescattering from charmless intermediate states
that receive sizable tree contributions. Most of the intermediate
states  such as $\ov K^0\eta',K^{*-}\rho^+,\cdots$, in $B$ decays
are penguin dominated and hence will not affect $S_f$. For the
decay $\ov B^0\to \phi K_S$, we have rescattering from $K^-\rho^+$
and $K^{*-}\pi^+$. Because of the absence of the penguin chiral
enhancement in $\ov B^0\to K^{*-}\pi^+$ and the large cancellation
between $a_4$ and $a_6$ penguin terms in $\ov B^0\to K^-\rho^+$,
it follows that the color-allowed tree contributions in these two
modes are either comparable to or slightly smaller than the
penguin effects. As for the $\omega K_S$ mode, there is only one
rescattering diagram, namely, $\ov B^0\to K^{*-}\pi^+\to \ov
K^0\omega$, arising from the charmless intermediate states. (The
rescattering diagram from $K^{*-}\rho^+$ is suppressed as
elucidated before for the $\phi K_S$ mode.) As a result, one will
expect that final-state rescattering effect on $S_f$ will be most
prominent in $B\to \omega K_S$, $\rho^0K_S$ and $\phi K_S$.
Indeed, we see from Table \ref{tab:CP} that FSI lowers $S_{\omega
K_S}$ by 15\% and enhances $S_{\rho^0K_S}$ by 17\% and $S_{\phi
K_S}$ slightly. The theoretical predictions and experimental
measurements for the differences between $S_f^{SD+LD}$ and
$S_{J/\psi K_S}$, $\Delta S_f^{SD+LD}$, are summarized in Table
\ref{tab:Sf}. It is evident that final-state interactions cannot
induce large $\Delta S_f$ in any of these modes.

\begin{table}[t]
\caption{Direct \CP asymmetry parameter $\A_f$ and the
mixing-induced \CP parameter $\Delta S_f^{SD+LD}$ for various
modes. The first and second theoretical errors correspond to the
SD and LD ones, respectively (see the text for details). The
$f_0K_S$ channel is not included as we cannot make reliable
estimate of FSI effects on this decay.} \label{tab:Sf}
\begin{ruledtabular}
\begin{tabular}{l r c c r c c}
 &  \multicolumn{3}{c}{$\Delta S_f$}
 &   \multicolumn{3}{c}{$\A_f(\%)$}  \\ \cline{2-4} \cline{5-7}
\raisebox{2.0ex}[0cm][0cm]{Final State} & SD & SD+LD & Expt & SD &
SD+LD & Expt
\\ \hline
 $\phi K_S$ & $0.02^{+0.00}_{-0.04}$ & $0.03^{+0.01+0.01}_{-0.04-0.01}$ & $-0.38\pm0.20$  &
  $1.4^{+0.3}_{-0.5}$ & $-2.6^{+0.8+0.0}_{-1.0-0.4}$  & $4\pm17$ \\
 $\omega K_S$ & $0.12^{+0.05}_{-0.06}$  & $0.01^{+0.02+0.02}_{-0.04-0.01}$  & $-0.17^{+0.30}_{-0.32}$
 & $-7.3^{+3.5}_{-2.6}$  & $-13.2^{+3.9+1.4}_{-2.8-1.4}$ & $48\pm25$ \\
 $\rho^0K_S$ & $-0.09^{+0.03}_{-0.07}$ & $0.04^{+0.09+0.08}_{-0.10-0.11}$ & -- & $9.0^{+2.2}_{-4.6}$ &
 $46.6^{+12.9+3.9}_{-13.7-2.6}$ & -- \\
 $\eta' K_S$ & $0.01^{+0.00}_{-0.04}$ & $0.00^{+0.00+0.00}_{-0.04-0.00}$
 & $-0.30\pm0.11$  & $1.8^{+0.4}_{-0.4}$ & $2.1^{+0.5+0.1}_{-0.2-0.1}$ & $4\pm8$ \\
 $\eta K_S$ & $0.07^{+0.02}_{-0.04}$ & $0.07^{+0.02+0.00}_{-0.05-0.00}$ & $-$
 & $-6.1^{+5.1}_{-2.0}$ & $-3.7^{+4.4+1.4}_{-1.8-2.4}$ & $-$ \\
 $\pi^0K_S$ & $0.06^{+0.02}_{-0.04}$ & $0.04^{+0.02+0.01}_{-0.03-0.01}$ & $-0.39^{+0.27}_{-0.29}$ &
 $-3.4^{+2.1}_{-1.1}$ & $3.7^{+3.1+1.0}_{-1.7-0.4}$ & $-8\pm14$  \\
\end{tabular}
\end{ruledtabular}
\end{table}

It is interesting to study the correlation between $\A_f$ and
$S_f$ for the penguin dominated modes in the presence of FSIs. It
follows from Eq. (\ref{eq:CfSf}) that
  \be
 {\Delta S_f\over \A_f}= \cos 2\beta\cot\delta_f\approx
 0.95\cot\delta_f,
 \label{eq:DSf/Cf}
 \en
for $r_f=(\lambda_u A_f^u)/(\lambda_c A_f^c)\ll 1$. This ratio is
independent of $|r_f|$ and hence it is less sensitive to hadronic
uncertainties. Therefore, it may provide a better test of the SM
even in the presence of FSIs. Writing
$A^{u,c}_f=|A^{u,c}_{SD}|e^{i\delta^{u,c}_{SD}}+i A^{u,c}_{LD}$,
we have
 \be
 |r_f|&=&\frac{|\lambda^u|}{|\lambda^c|}
 \frac{\sqrt{|A^u_{SD}\cos\delta^u_{SD}|^2+(|A^u_{SD}|\sin\delta^u_{SD}+A^u_{LD})^2}}
 {\sqrt{|A^c_{SD}\cos\delta^c_{SD}|^2+(|A^c_{SD}|\sin\delta^c_{SD}+A^c_{LD})^2}},
 \non\\
 \delta_f&=&\tan^{-1}(\tan\delta^u_{SD}+\frac{A^u_{LD}}{|A^u_{SD}|\cos\delta^u_{SD}})
       -\tan^{-1}(\tan\delta^c_{SD}+\frac{A^c_{LD}}{|A^c_{SD}|\cos\delta^c_{SD}}),
 \en
where the reality of $A_{LD}$ has been used, and $\A_f$ and
$\Delta S_f$ can be obtained by using Eq.~(\ref{eq:CfSf}). It is
interesting to note that in the absence of LD contributions we
have $|\delta_f^{SD}|=|\delta^u_{SD}-\delta^c_{SD}|\lsim\pi/4$ for
some typical SD (perturbative) strong phases and, consequently, we
expect $|\Delta S_f/\A_f|\gsim 1$. This result generally does not
hold in the presence of FSI. For example, in the case of
$|A^u_{LD}|\ll |A^{u,c}_{SD}|\lsim|A^c_{LD}|$, it is possible to
have $|\Delta S_f/\A_f|\lsim 1$. From Table \ref{tab:Sf} we obtain
 \be \label{eq:S/C}
 && \Delta S_{\phi K_S}/\A_{\phi K_S}\approx -1.3\,(1.5), \quad \Delta S_{\omega K_S}/
 \A_{\omega K_S}\approx -0.08\,(-1.7), \quad \Delta S_{\rho^0 K_S}/
 \A_{\rho^0 K_S}\approx 0.08\,(-1.1), \non \\
 && \Delta S_{\eta' K_S}/\A_{\eta' K_S}\approx -0.05\,(0.6), \qquad
 \Delta S_{\eta K_S}/\A_{\eta K_S}\approx -2.0\,(-1.1), \qquad \Delta S_{\pi^0 K_S}/
 \A_{\pi^0 K_S}\approx 1.2\,(-1.8), \non \\
 \en
where the SD contributions are shown in parentheses. For the
$f_0K_S$ mode, $\Delta S/\A\approx 3.0$ but there is no reliable
estimate of the FSI effect. From Eq. (\ref{eq:S/C}) we see that
the relation $|\Delta S_f/\A_f|\gsim 1$ indeed holds at short
distances except for the $\eta'K_S$ mode where
$|\delta_f^{SD}|\sim 65^\circ$. The sign flip of $\Delta S_f/\A_f$
in the presence of LD rescattering is due to the sign switching of
either $\A_f$ (for $\phi K_S$ and $\pi^0K_S$) or $\Delta S_f$ (for
$\rho^0K_S$).

\section{Conclusions and Comments}
In the present work we have studied final-state rescattering
effects on the decay rates and \CP violation in the
penguin-dominated decays $B^0\to
(\phi,\omega,\rho^0,\eta',\eta,\pi^0,f_0)K_S$. Our main goal is to
understand to what extent indications of possibly large deviations
of the mixing-induced \CP violation seen in above modes from $\sin
2\beta$ determined from $B\to J/\psi K_S$ can be accounted for by
final-state interactions. Our main results are as follows:

\begin{enumerate}

\item We have applied the QCD factorization approach to study the
short-distance contributions to the above-mentioned seven modes.
There are consistently 2 to 3 $\sigma$ deviations between the
central values of the QCDF predictions  and the experimental data.

\item The differences between the \CP asymmetry $S^{SD}_f$ induced
at short distances and the measured $S_{J/\psi K_S}$ are
summarized in Eq. (\ref{eq:Ssd}). The deviation of $S_f^{SD}$ in
the $\omega K_S$ and $\rho^0K_S$ modes from $\sin 2 \beta$ is a 2
to 3 $\sigma$ effect owing to a large tree pollution. In contrast,
tree pollution in $\eta' K_S$ is diluted by the QCD anomaly and/or
final-state rescattering both of which carry negligible $CP$-odd
phase. The long-distance effects on $S_f$ are generally negligible
except for the $\omega K_S$ and $\rho^0K_S$ modes where $S_f$ is
lowered by around 15\% for the former and enhanced by the same
percentage for the latter and $\Delta S_{\omega
K_S,\rho^0K_S}^{SD+LD}$ become consistent with zero within errors.

\item Final state rescattering effects from charm intermediate
states can account for the discrepancy between theory and
experiment for the branching ratios of the modes $\omega
K_S,\eta'K_S,\phi K_S,$ and $\pi^0K_S$.  Moreover, direct \CP
asymmetries in these modes are significantly affected; the signs
of $\A_f$ in the last two modes are flipped by final-state
interactions. Direct \CP asymmetries in the $\omega K_S$ and
$\rho^0K_S$ channels are predicted to be $\A_{\omega K_S}\approx
-0.13$ and $\A_{\rho^0K_S}\approx 0.47$, respectively, which
should be tested experimentally.

\item For the $f_0(980)K_S$ mode, the short-distance contribution
gives $\Delta S/\A\approx 3.0$, but at present we cannot make
reliable estimates of FSI effects on this channel.

\item  Direct \CP asymmetry in all the $(b\to s)$
penguin-dominated modes is rather small ($\lsim$ a few \%) except
for $\omega K_S$ and $\rho^0K_S$. This strengthens the general
expectation that experimental search for direct \CP violation in
$b\to s$ modes may be a good way to look for possible effects of
new $CP$-odd phases (see e.g. \cite{AS97}).

\item Since the mixing induced \CP parameter $S_f$ (actually
$\Delta S_f\equiv -\eta_fS_f-S_{J/\psi K_S}$) and the direct \CP
parameter $\A_f$ are closely related, so are their theoretical
uncertainties. Based on this study, it seems rather difficult to
accommodate $|\Delta S_f| > 0.10$ within the SM, at least in the
modes we study in this paper (except for $f_0K_S$ which we cannot
make reliable estimates).

\item In particular, $\eta'K_S$ and (to some degree) $\phi K_S$
appear theoretically cleanest in our picture; i.e. for these modes
the central value of $\Delta S_f$ as well as the uncertainties on
it are rather small. This also seems to be the case in QCDF
\cite{Beneke-ckm05}. Note also that the experimental errors on
$\eta'K_S$ are the smallest (see Table III) and its branching
ratio is the largest, making it especially suitable for faster
experimental progress in the near future \cite{Smith}.

\item The sign of $\Delta S_f$ at short distances is found to be
positive except for the channel $\rho^0K_S$. After including
final-state rescattering effects, the central values of $\Delta
S_f$ are positive for all the modes under consideration, but they
tend to be rather small compared to the stated uncertainties so
that it is difficult to make reliable statements on the sign at
present. However, since the $S_f$ and $\A_f$ are strongly
correlated, improved measurements could provide enough useful
information that stronger statements on the sign could be made in
the future.

\end{enumerate}

\vskip 2.0cm \acknowledgments
 We thank Martin Beneke, Tom Browder, Masashi Hazumi, Luca Silvestrini
and Jim Smith for discussions. This research was supported in part
by the National Science Council of R.O.C. under Grant Nos.
NSC93-2112-M-001-043, NSC93-2112-M-001-053 and by the U.S. DOE
contract No. DE-AC02-98CH10886(BNL).


\end{document}